\documentclass[draft,jgrga]{AGUTeX}

\usepackage{amsmath}
\usepackage{amsbsy}
\usepackage{rotating}
\usepackage{amsfonts} 

\setkeys{Gin}{draft=false}

\authorrunninghead{MULLIGAN ET AL.}

\titlerunninghead{Advancing In Situ Modeling of ICMEs}

\begin{document}


%


%


\title{Advancing In Situ Modeling of ICMEs: New Techniques for New Observations}

%



%


%


%





\authors{T. Mulligan,\altaffilmark{1}, Alysha A. Reinard\altaffilmark{2}, and
        Benjamin J. Lynch\altaffilmark{3}}

\altaffiltext{1}{Space Sciences Department, The Aerospace Corporation,
Los Angeles 90009 USA; tamitha.mulligan@aero.org}

\altaffiltext{2}{NOAA Space Weather Prediction Center, Boulder, CO 80505, USA; alysha.reinard@noaa.gov}

\altaffiltext{3}{Space Sciences Laboratory, University of California,
        Berkeley, CA 94720, USA; blynch@ssl.berkeley.edu}








%


%




\begin{abstract}

It is generally known that multi-spacecraft observations of interplanetary coronal mass ejections (ICMEs) more clearly reveal their three-dimensional structure than do observations made by a single spacecraft. The launch of the STEREO twin observatories in October 2006 has greatly increased the number of multipoint studies of ICMEs in the literature, but this field is still in its infancy. To date, most studies continue to use on flux rope models that rely on single track observations through a vast, multi-faceted structure, which oversimplifies the problem and often hinders interpretation of the large-scale geometry, especially for cases in which one spacecraft observes a flux rope, while another does not. In order to tackle these complex problems, new modeling techniques are required. We describe these new techniques and analyze two ICMEs observed at the twin STEREO spacecraft on 22-23 May 2007, when the spacecraft were separated by $\sim$8$^{\circ}$. We find a combination of non-force-free flux rope multi-spacecraft modeling, together with a new non-flux rope ICME plasma flow deflection model, better constrains the large-scale structure of these ICMEs.  We also introduce a new spatial mapping technique that allows us to put multispacecraft observations and the new ICME model results in context with the convecting solar wind. What is distinctly different about this analysis is that it reveals aspects of ICME geometry and dynamics in a far more visually intuitive way than previously accomplished. In the case of the 22-23 May  ICMEs, the analysis facilitates a more physical understanding of ICME large-scale structure, the location and geometry of flux rope sub-structures within these ICMEs, and their dynamic interaction with the ambient solar wind.

\end{abstract}


%


%



%



\begin{article}

\section{Introduction}

Coronal mass ejections (CMEs) are eruptions from the Sun that release vast quantities of plasma into the expanding solar wind. When observed at 1 AU, the interplanetary counterpart of CMEs (ICMEs) often exhibit depressed plasma temperatures, enhanced and twisted magnetic fields, and bi-directional suprathermal electron signatures making them distinguishable from the ambient solar wind [Burlaga et al., 1981; Burlaga and Behannon, 1982, Gosling, 1990; Neugebauer and Goldstein, 1997; Richardson and Cane, 2004a; Zurbuchen and Richardson, 2006)]. In the interplanetary medium, the appearance of ICMEs varies greatly such that only about one-third to one-half of all ICMEs in the inner heliosphere are observed as magnetic clouds (MCs) (Gosling 1990; Wang and Richardson, 2004), though this may be a product of viewing geometry, as recent results suggest ICMEs intersected far from their axis do not exhibit MC signatures (Jian, 2006, Kilpua, 2009, 2011). In MCs, simplifying assumptions have been made about their three-dimensional structure and they have successfully been modeled as magnetic flux ropes using force-free bessel functions based on the Lundquist solution (Lundquist 1950; Goldstein, 1983; Burlaga et al., 1990, Kilpua et al., 2011 and references therein). This basic flux rope model assumes that ICMEs are in equilibrium and have circular cross-sections. However, in-situ studies with multiple spacecraft have found ICMEs are dynamic structures that undergo extensive expansion and interact with the ambient solar wind. These processes result in mild to extreme distortions in the magnetic field geometry and flux rope cross-sectional shapes (e.g., Gosling, 1990; Farrugia et al., 1995; Mulligan et al., 1999; Mulligan and Russell, 2001; Russell and Mulligan, 2002; Riley and Crooker, 2004; Lepping et al., 2006; Liu et al., 2006). Consequently, the Lundquist flux rope is now considered an approximate solution and has prompted several revisions to the basic model, such as the inclusion of non-constant alpha and expansion effects (Marubashi, 1986, 1997), uniform-twist (Farrugia et al., 1999), and non-force free kinematic effects (Mulligan and Russell, 2001; Cid et al., 2002; Hidalgo et al., 2002, Owens et al., 2006).

Although flux rope modeling has dominated ICME analyses over the past decade, there is a significant fraction of ICMEs that have no observable flux rope signature, which fuels the debate as to whether or not an ICME must contain a flux rope in its interior.   In the ecliptic plane ICMEs typically span up to 60$^{\circ}$ in longitude (consistent with the angle the average CME subtends at the Sun) so there are many examples in which multiple spacecraft have observed the same ICME. However, the probability of intercepting a central flux rope structure within the ICME diminishes with increasing impact parameter (the closest approach to the center of the ICME). In the Jian et al., (2006) study, a majority of ICME observations occur with the observing spacecraft intercepting the ICME disturbance sufficiently far from its center such that a central flux rope is not identifiable.   Kilpua et al., (2009, 2011) also show examples of multispacecraft observations in which one spacecraft crosses the central flux rope near its apex, while the other spacecraft traverses the ICME flank, where the flux rope structure is no longer discernible.Thus at 1 AU, the association between multiple spacecraft observations of an ICME is not always straightforward, resulting in significant differences, even when separation distances of the spacecraft are small. In addition, ICME characteristics such as size, magnetic field and plasma signatures, and the structure of the surrounding solar wind can vary significantly from event to event. Despite this difficulty, multi-spacecraft observations, separated by at least a few degrees, comprise the best tool currently available to study the large-scale properties of these vast structures. Since the majority of ICME observations either come from single spacecraft or multispacecraft encounters that are difficult to reconcile, the three dimensional structure of ICMEs is still poorly understood. 

Data from the Solar TErrestrial Relation Observatory (STEREO) (Kaiser et al., 2007), launched in 2006, are ideally suited to help characterize the elusive, large-scale properties of ICMEs. STEREO consists of two functionally identical satellites, one that leads the Earth (STEREO-A), and one that lags the Earth (STEREO-B) in its orbit around the Sun with gradually increasing angular separation. The focus of this paper is to better exploit the available multi-spacecraft observations  by combining measurements from STEREO with those from near-earth spacecraft like Wind and ACE (approximately halfway between the STEREO spacecraft) into a single, coherent model reconstruction of large-scale ICME strucutre.   Promising methods already exist, such as those based on the Grad-Shafranov Reconstruction (GSR) technique, which physically constrain magnetic field and plasma observations into a single model inversion (Hu and Sonnerup, 2001, 2002). However,  models based on this technique have only been applied to flux rope observations. Though GSR-based models have merits, in practice they also also suffer from non-uniqueness and have been shown to incorrectly determine the shape of flux rope cross-sections in complex MHD simulations of ICMEs (e.g. Riley 2004). 

To advance beyond these existing reconstructions, new methods must be employed to tackle complex ICME structures, such as those having no discernible magnetic flux rope signature at one or more spacecraft locations.  Ideally, such a model would include magnetic field and plasma measurements of the surrounding solar wind to constrain the inversion and to help characterize the distortion and dynamics of ICME structure on a large scale. Consequences of ICMEs are far-reaching in the heliosphere and there is a need to create a more realistic three-dimensional  picture of these structures in context with the ambient interplanetary environment. Utilizing data in this way gives us clues to how ICMEs evolve in the solar wind and what relationship they share with the surrounding plasma envelope, its composition, sub-structures, and dynamics. Overall, it is this better understanding of the large-scale structure that is of key importance for solar-terrestrial research and space weather forecasting.

In the following sections we outline the development of two new techniques designed to extend the utility of the Mulligan and Russell (2001) magnetic flux rope model to simultaneously invert data from flux rope and non-flux rope multi-spacecraft ICME observations. Specifically, in Section 2 we outline a technique that uses the solar wind plasma velocity data and the ICME boundary normal to allow the inclusion of non-flux-rope observations in a new large-scale or “synoptic” ICME model inversion. In Section 3 we discuss the development of a new spatial mapping technique that exploits solar wind magnetic field, plasma, and ionic composition measurements observed at multiple spacecraft in order to create a more quantitative understanding of the ambient plasma environment and how this environment affects the dynamics and evolution of the ICME structure. Also in this section (and in the Appendix) we test the spatial mapping method on a period of solar wind, when the ACE, Wind, and twin STEREO spacecraft are in close proximity and reconstruct the time-series data from the spatial maps.  In Section 4 we apply the new techniques developed in Sections 2 and 3 to multispacecraft measurements of two ICMEs on 22-23 May 2007 seen at STEREO-A, STEREO-B, ACE, and Wind. In the closing sections 5 through 7, we discuss the implications of this work and future developments that promise to usher in a new era in the analysis of multi-spacecraft observations.

\section{Development of the Synoptic ICME model }

As noted previously, MCs are well approximated by flux rope models, but they are not always observable when an ICME is intersected far from its central axis.  When this occurs, magnetic field observations do not provide enough information about the global ICME structure at this location. However, velocity flow deflections around the ICME can provide the missing insight into the large-scale orientation of the structure in these cases and by incorporating this information into an existing flux rope model we can extend its capabilities.

The baseline for the development of the synoptic ICME model is the non-cylindrically symmetric, non-force-free flux rope model described in detail in Mulligan and Russell, (2001) and Mulligan (2002). Briefly explained, this model has an axial magnetic field component that falls off with a stretched-exponential form from the axis of the rope. The azimuthal or poloidal field increases as 1 minus a stretched-exponential dependence so that it maximizes at the rope edge. The stretched exponential form allows the modeling of non-cylindrically symmetric cross sectional geometries. When used with multiple spacecraft observations the model returns the azimuthal stretching of the flux rope cross-section, the bending along the flux rope axis, and the spatial (temporal) expansion of the modeled structure. The model flux rope is fit to the data using a downhill simplex inversion technique (Nelder and Mead, 1965) that varies the fitting parameters in an orderly manner.  The parameters already allow for gross measurements of cross-sectional distortion, large-scale curvature, and residual expansion forces, and by adding velocity component modeling, these quantities will be better resolved. 

 In general, velocity flow deflections are only an approximation of the ICME orientation, subject to the assumptions of obstacle shape, and are not sufficient alone to determine ICME structure, especially from single spacecraft observations.  However, when inverting multiple observations, for example, in which one spacecraft observation does not have a clear flux rope signature, the addition of this data can provide a valuable constraint on the large-scale geometry of the entire structure and improve the overall fit dramatically. 

\subsection{Velocity component modeling (Lindsay technique)}  

 The first step towards incorporating solar wind flow deflection around an ICME obstacle originates in the work by G. Lindsay, (1996). In her thesis she exploited the fact that as an ICME travels outward faster than the solar wind ahead, the solar wind ahead must be deflected away from and around the ICME. This implies that non-radial, solar wind velocity components will be associated with ICME passage.  For clarity, it should be emphasized that it is the deflection across the ICME leading boundary that is being examined, not the deflections across the interplanetary shock associated with the ICME. This means that slow ICMEs, which do not have shocks can also be studied using this method, provided they are moving faster than the ambient solar wind.  This effect has been previously studied by Gosling et al., (1987) for east-west flow deflections in ICMEs at 1 AU. By considering east-west and north-south flow deflections and making the assumption that the gross structure of an ICME is cylindrical, Lindsay et al., (1996) was able to determine the relationship between the velocity flow deflection around the ICME obstacle and the obstacle orientation. The can diagrams in Figure 1 illustrate how this determination is performed. For an ICME with its axis in the ecliptic plane (Figure 1a), the solar wind velocity will be deflected both above and below the ICME obstacle as it passes through the upstream solar wind. This means that only northward or southward velocity deflections (in GSE coordinates, deflections in $v_z$) will be observed ahead of the ICME.  In the other extreme, an ICME with its axis oriented perpendicular to the ecliptic plane (Figure 1b) will produce only eastward or westward deflections (deflections in $v_y$) ahead of the ICME.

Assuming that an ICME can be approximated by a cylindrical structure, the bottom diagrams in Figure 1 illustrate the expected flow deflections of the velocity components as the spacecraft traverses the ICME via trajectory T1 or T2. As an example, for an observing trajectory following that of T1 in Figure 1b, sufficiently upstream of the ICME (where the influence of the ICME is not present) radially oriented solar wind flow will be observed. As the ICME approaches the observation location, the presence of the obstacle will begin to influence the solar wind flow and the speed of the solar wind plasma will begin to increase.  Within the region of increasing plasma speed and across the ICME leading boundary (at time $t_0$ in the figure), the plasma will be observed to be deflected eastward ($v_y>0$). Inside the ICME, a westward deflection ($v_y<0$) will be observed.  Through the remainder of the ICME, the solar wind speed begins to decline until, near the trailing boundary (at time $t_1$), it attains a radial orientation. If the observing spacecraft follows trajectory T2, the observed $v_y$ deflections will occur in the opposite sense (i.e.  westward, then eastward). These deflections show a predictable pattern at the ICME leading edge boundary and reveal a circulation pattern within the ICME structure (Lindsay 1996). 

The systematic variations in ICME related velocity flow deflections can be used to infer ICME orientation in the plane of the sky. Again, for example, the $v_y<0$ deflection ahead of the ICME shown in Figure 1b indicates that the spacecraft is intercepting the ICME in a manner similar to trajectory T2. For this same trajectory, a flow deflection $v_y<0$ and $v_z>0$ ahead of the ICME implies that the ICME is oriented as in Figure 2a. Formally, the orientation of the cylinder axis can be calculated from the observed velocity flow deflections by
\begin{equation}
tan\theta_c = - \frac{v_z}{v_y}
\end{equation}
where $\theta_c$ is defined as in Figure 2a and where $\theta_c > 0$ indicates an eastward orientation and  $\theta_c < 0$ represents a westward orientation.  In the extremes that $v_z=0$ or $v_y=0$, implies that the ICME is either perpendicular to or aligned with the ecliptic plane, respectively. 

Lindsay (1996) performed this analysis on 23 MCs observed by the Pioneer Venus Orbiter spacecraft and found orientations in the plane of the sky (a.k.a. the clock angle, defined in the y-z plane in GSE coordinates as the angle with 0$^{\circ}$ pointing due northward and a positive angle in the positive y direction in a counterclockwise sense) that were consistent with the Lundquist flux rope model solutions performed by Lepping (1990).  The clear advantage of the Lindsay method, especially with respect to ICME modeling, is that it can be used on spacecraft traversals of non-flux rope ICMEs and traversals of suspected flux rope ICMEs for which there is little or no magnetic field information available.

\subsection{Velocity component modeling (Owens and Cargill technique)}

Because the Lindsay technique (Lindsay 1996) only determines ICME orientation in the plane of the sky (a.k.a. the clock angle), additional information must be acquired to determine the obstacle orientation out of this plane. Using a method outlined by Owens and Cargill, (2004), it is possible to expound upon the Lindsay technique and make a quantitative comparison of the complete flux rope model orientation by calculating the normal to the ICME obstacle using velocity flow deflections.  

Figure 2 summarizes how the flow deflections from the Lindsay technique are used to determine the obstacle clock angle. In the work by Owens and Cargill (2004), two unit vectors in the frame of the ICME must be determined to obtain the normal to the ICME obstacle.  Retaining the naming convention for these vectors as in the Owens and Cargill paper, the unit vector $\hat{a}$ can be expressed as a function of the components of the velocity flow deflection in the y-z plane
\begin{equation}
\hat{a} = \frac{v_y}{|v|}\hat{y} - \frac{v_z}{|v|}\hat{z}
\end{equation}

where $\sqrt{v_y^2 + v_z^2}$. Recall that for a cylindrical obstacle oriented with its axis perpendicular to the ecliptic plane, a vector tangent to the obstacle at the spacecraft location is the velocity flow deflection in the x-y plane. Figure 2b shows the cross section of a cylindrical obstacle oriented with its axis perpendicular to the ecliptic plane.  The unit vector $\hat{c}$ is tangent to the obstacle surface and points in the direction of the velocity flow deflection in the x-y plane at point p. Note that $\hat{c}$ can be related to the impact parameter (IP) of the spacecraft trajectory through the obstacle by the angle $\alpha$, where $\alpha$ is the angle between the x-axis of the rope and the radial line between the center of the cross-section and the spacecraft point of entry, as Figure 2b shows. Thus it is possible to calculate the impact parameter using the flow deflection components in the x-y plane 
\begin{equation}
IP = sin(tan^{-1}(-\frac{v_y}{v_x})).
\end{equation}
As first suggested by Owens and Cargill, by crossing vector $\hat{a}$  with  $\hat{c}$, the obstacle normal $\hat{n}$  can be determined. This is illustrated in Figure 2c for two different impact parameters.

For simplicity, the diagrams in Figure 2 do not consider all possible orientations of the ICME obstacle (i.e. they ignore axial orientations out of the plane of the page), nor do they illustrate all non-cylindrical rope cross-sections (Mulligan and Russell 2001). However, the methods outlined can be easily generalized to any rope orientation and elliptical cross-section.  In this fashion, both the normal obtained from the additional spacecraft data and the normal from the magnetic field model can be compared and the residuals between these normals minimized. The additional impact parameter information helps to constrain the geometry of the ICME relative to the spacecraft location. In this sense, the minimization of the normals becomes part of the free parameter space of the synoptic model and a new, higher-dimensional response surface is created. Due to the versatility of the Nelder and Mead simplex algorithm (Nelder and Mead, 1965), this new response surface is easily traversed in the optimization process.

It is worthy to note that the determination of the spacecraft impact parameter and obstacle normal (based on the flow deflection components in the x-y plane) is heavily dependent upon the choice of the reference frame, which must be co-moving with the ICME obstacle, such that the bulk flow of the solar wind (primarily in the radial direction) is minimized.  One way this can be performed is by transforming into the deHoffman-Teller frame in which a velocity V$_{HT}$ is found from the reference frame to the spacecraft frame that it minimizes the residual electric field in the least-squares sense (de Hoffmann and Teller, 1950). Other methods for determining the appropriate velocity of the co-moving reference frame can also be applied and include the Constant Velocity Approach (CVA; Russell et al., 1983), the Minimum Faraday Residue (MFR; Terasawa et al., 1996; Khrabrov and Sonnerup 1998) and Minimum Mass-Flux-Residue (MMR; Sonnerup et al., 2004), which will not be described here. At the current time, we employ a simple approach of transforming into a frame co-moving with the average radial solar wind speed of the ICME. 

Additional methods to determine the ICME normal such as the coplanarity theorem can be applied successfully when magnetic field data is available, the variance of the field remains low, and when the magnetic field direction is not aligned parallel or perpendicular to the obstacle normal. If no useful magnetic field data exists,  the velocity coplanarity normal is an alternate method. The mixed-mode normal, which employs both the field components and the velocity components in the spacecraft frame (Abraham-Shrauner and Yun, 1976; Russell et al., 1983) is another possibility. The advantage of using these other methods for calculation of the obstacle normal is that they do not suffer from errors caused by a translation into a co-moving frame.  In a future publication we plan to refine the technique by computing the ICME normal using the various methods mentioned above and average the result over an ensemble of such pairs.  This process will build-up an ensemble statistical deviation and give an error estimate of the result. (e.g. Russell et al., 1983; Schwartz (1998)).

\section{Spatial mapping technique}

Spatial interpolation of data between spacecraft is not new to analyses of space plasmas. The advent of multi-spacecraft missions has resulted in the development of several techniques for  multipoint data analyses (e.g. Hu and Sonnerup, 2002; Chanteur and Mottez, 1993). In particular, Chanteur (1998), outlines the use of barycentric coordinates and reciprocal vectors, a method well-known in applied mathematics and computational geometry, but not yet fully explored in space physics. Most recently this technique has been used to interpolate data within the four-spacecraft Cluster tetrahedron to infer three-dimensional spatial information about transient structures in the Earth’s magnetosphere.  The technique we introduce in this section employs a similar method as that from Chanteur and Mottez, (1993) using barycentric coordinates, but in our case we limit the problem to a two-dimensional geometry.  This allows us to simplify the interpolation algorithm using a Delaunay triangulation of the data on a two-dimensional spatial grid. Background and mathematical details of Delaunay triangulation using the natural neighbor interpolation method (Sibson, 1981) are found in the Appendix. The following sections discuss how we transform time-series data onto a spatial domain and apply this Delaunay-Sibson interpolation method.

\subsection{Defining the Spatial Domain}

In the case of multiple spacecraft separated by several degrees in longitude in the ecliptic plane, the relatively small latitudinal differences between the spacecraft can be ignored, reducing the system to a two-dimensional space in which temporal data can be mapped into the spatial domain. This mapping can be accomplished by a linear transformation of the time-series data using the bulk plasma velocity of the solar wind at each time step.  Using the spacecraft locations at time $t_o$ as the midpoint of the spatial grid, we can track the spacecraft positions in the spatial domain at n earlier (t$_o$-n$\Delta$t) and n later (t$_o$+n$\Delta$t) times using 
\begin{equation}
(r_{n^+} - r_0) = \sum_{k=1}^{n} v_k \cdot \Delta t
\end{equation}
\begin{center} and \end{center}
\begin{equation}
(r_{n^-} - r_0) = \sum_{k=1}^{n} -v_k \cdot \Delta t
\end{equation}
where $\Delta$t indicates the time resolution of the data, $r_o$ is the position vector at the midpoint of the spatial grid, $v_k$ is the vector measurement of the solar wind bulk velocity at time k, and 2n is the number of data samples in the time series. The mapped data is centered on time t$_o$ to minimize propagation errors from the velocity data in the spatial range calculation. Figure 3 highlights the use of this technique in two-dimensions (R, T) in which R is the radial dimension and T is the transverse dimension.  Note the sampling frequency will affect the resolution of features in the map. (For more details about sampling frequency and error  see the discussion in the Appendix.) 

Once time-series for each spacecraft has been mapped as separate spacecraft tracks through the spatial domain, we perform a Delaunay triangulation, which computes a set of two-dimensional simplices (triangles) from the points comprising the spacecraft tracks in the spatail domain (a tessellation, e.g. Figure A5). Each point in the tessellation has a spatial location as well as a corresponding data value for the observed solar wind parameters (e.g. $Np$, $|Vp|$, $Tp$, and $B$). These values are then interpolated between the spacecraft tracks using the weighted estimators determined from the Sibson natural neighbor method (Sibson, 1981; e.g. Figure A4).  By interpolating over the multi-spacecraft data set in the entire spatial domain, scalar and vector fields of the solar wind parameters can be created. Because the spatial map and the weighted estimators are solely a function of the Delaunay triangulation geometry, the spatial grid is not recalculated for each of the solar wind parameters. Thus, for example, the solar wind $Np$, $|Vp|$, and $Tp$ scalar fields shown in Figure 4 are calculated using the same underlying Delaunay triangulation and weighted estimators.

Because the spacecraft "trajectories" in the maps are created from the solar wind convecting outward past near-stationary spacecraft locations, the mapping is not analogous to a snapshot in time at all locations (e.g. at all R, and T, from Figure 3). A resultant spatial map is more akin to a running time plot, in which snapshots at a single time and location are captured, propagated forward, and pieced together.  Since the solar wind moves at a finite speed, the spacecraft trajectories in the spatial domain take a finite amount of time to traverse the grid (equivalent to the time it takes to measure the 2n samples in the time series). This results in an inherent time dependence in the maps, such that any interpolated data nearest to the left side of the plot (e.g. in Figure 3) will have a $2n\Delta t$ difference in time than those at the right hand side of the plot.

\subsection{Spatial mapping test case: March 25-30, 2007 }

To illustrate the usefulness of this technique and perform a preliminary error analysis, we perform a Delaunay triangulation and Sibson interpolation of ACE, Wind, STEREO-A (STA), and STEREO-B (STB) plasma data during a moderately disturbed period centered on 27 March 2007, when the STEREO spacecraft separation was ~3.6$^{\circ}$ (approximatley half the separation distance during the 22-23 May 2007 event series analyzed in Section 4).  In this test case, 10-minute averaged data is used. Contour maps resulting from the spatial mapping and interpolation processes are shown in Figure 4 for the proton bulk speed $|Vp|$. (Spatial maps for the other bulk plasma parameters of the solar wind, $Np$, and $Tp$ are shown in the Appendix.) In each panel the time series data ha been converted to spatial coordinates, with the origin of the spatial map located at the Earth's position in the y direction and at the right  side of the figure in the x direction. The axes are given in AU in the ecliptic plane. 

We chose this interval because it allows a test of how well the interpolation can reconstruct observed data between two spacecraft separated by 3.6$^{\circ}$, which is the mean separation distance between STA and ACE in Figure 9. We purposely use a time period of nearly five days, with a  time $t_o$ defined at the beginning of the time series to maximize the effect of solar wind velocity error propagation. Four separate Delaunay triangulations and Sibson interpolations are performed between the various spacecraft data. In the top panel we use data from all four spacecraft (STA, ACE, Wind, and STB) against which we compare the other less constrained maps. The middle two panels use data from three spacecraft (STA, Wind, and STB followed by STA, ACE, and STB).  The bottom panel uses data from only STA and STB in the mapping and interpolation processes.  Grey lines in each panel show which spacecraft data tracks are used in the interpolation and how the spacecraft trajectories track through the spatial domain. As made obvious from viewing the growing separation between the ACE and Wind tracks in the y (transverse) dimension, from the right of the plot (e.g. beginning of the data series) to the left of the plot (e.g. end of the data series), there is a transverse distortion of the map due to propagation of error in the velocity data. The amount of transverse distortion between ACE and Wind is on the same order as the y-distance  (transverse spread) between them.  Despite this limitation, it is clear that the three-spacecraft interpolations (panels 2 and 3) and the two-spacecraft interpolation (panel 4) suffer only slight degradation from the four-spacecraft interpolation in the top panel. Upon close inspection, some fine-scale structures are either severely smoothed or completely lost in the three-spacecraft and two-spacecraft interpolations. However, fine-scale structures in the velocity exceeding ~0.03 AU in radial thickness are retained even in the map using data from only the two STEREO spacecraft.

By extracting interpolated data in the three- and two-spacecraft spatial maps corresponding to the missing ACE and Wind spacecraft tracks (missing grey lines in panels 2-4 of Figure 4), it is possible to more directly compare the results of the three- and two-spacecraft interpolations to the actual time-series data. This is accomplished through a reverse transformation of the interpolated data back to the time domain using the same bulk plasma velocity components in the ecliptic plane.  The purpose of this exercise is to determine to what degree the mapping and interpolation processes can recover spatial information between spacecraft data tracks and how well the observed time-series data can be reconstructed.   Figure 5 shows three examples of reconstructed solar wind $|V_p|$  $N_p$, and $T_p$ data for Wind. The top panel shows the comparison of Wind data for the contour maps of $|V_p|$ shown in Figure 4. In the plot, the blue line shows the observed Wind data, while the red and  black lines show the reconstructions from the three-spacecraft spatial map and two-spacecraft spatial map in Figure 4b and 4d, respectively.  Note the similarity of the interpolated data to the actual observations. 

The second and third panels show similar results for the reconstructed $N_p$ and $T_p$ data at Wind. The spatial maps associated with these two- and three-spacecraft time-series reconstructions are shown in Figures A5 and A6.  Note that the error is largest for the more highly variable $N_p$ and $T_p$ and lowest for the smoother profile of $|V_p|$. As expected, timing errors and amplitude errors also increase as the data moves further to the left of the map, away from  time t$_o$ (right side of the map). This is  due to the propagation of velocity error in the estimation of spatial grid (see Figure 3) and in practice the spatial mapping technique is not recommended for use over such long time intervals.

Overall, the success of the mapping and interpolation processes to reconstruct nearly quantitatively accurate time-series in a moderately disturbed period  lends confidence that the method is robust enough to apply to the 22-23 May 2007 event period. During the ICME event series, the spacecraft separation distances between the Earth positioned spacecraft (ACE and Wind) and the STEREO spacecraft will be on the same order as the separation between STA and STB in the 25-30 March 2007 exercise. Thus, spatial mapping of the 22-23 May 2007 event period should allow assessments of the bulk plasma parameters down to fine structures having thicknesses of $\sim$0.03 AU in the radial dimension.

\section{Observations}

On 19 May 2007, at 1324 UT, LASCO observed a halo CME originating from AR10956 and associated with a B9.5 flare. A second partial halo CME was launched less than a day later on 20 May 2007 from the same active region associated with a second B-class flare. The source region of these CME events has been well studied by Li et al., (2008) and references therein. The first CME was associated with a flux rope observed at STB and at L1, whereas clear signatures of the second CME were observed only by STA and at L1. On 21- 23 May 2007, the corresponding ICMEs arrived at ACE, Wind, STA, and STB, followed shortly thereafter by a high speed stream. STA and STB were separated by ~8.5$^{\circ}$. Additional details of the interplanetary observations can be found in Kilpua et al., (2009). 

Figures 6 and 7 show magnetic field and plasma time series observations from ACE, Wind, and the STEREO spacecraft during the ICME event series. In each figure, the top four panels show the magnetic field components in GSE coordinates for ACE and Wind and in an analogous coordinate system for STA and STB, in which the x direction points from the spacecraft towards the Sun, the z direction points northward, and the y direction completes the right-handed coordinate system. For consistency, we use the nomenclature and color coding from Kilpua et al., (2009) in labeling the two ICME regions MC1 and MC2 in each figure. Focusing ont Figure 6, both ACE and Wind observe the two ICMEs, MC1 and MC2 highlighted by the orange and blue regions, rspectively. Vertical dashed lines mark the leading and trailing boundaries of both ICMEs and indicate $t_o$ and $t_1$, the entry and exit velocity flow deflections (see Figure 1), which are particularly distinct in the $v_y$ and $v_z$ components.  The ICME modeling results (discussed in Section 4.1) are overlaid on the magnetic field components. Figure 7 shows similar time series data for STA and STB. In this figure, the MC1 region is split into two sections because of the different observing conditions at STA and STB.  The orange highlighted region in the top four panels mark the observation of MC1 at STB, with the magnetic flux rope fit overlaid on the magnetic field components.  Although an ICME signature was present, no flux rope was observed at STA for this event. Thus we have not highlighted any portion of the ICME magnetic field at STA for MC1. Instead, we highlight the velocity flow defection boundaries (between $t_o$ and $t_1$) at STA in the bottom three panels of Figure 7. Note that the ICME corresponding to MC1 is observed approximately four hours earlier at STA than  STB. Turning our focus to the second event in the series, MC2, which is only observed at STA, the blue region highlights both the magnetic field model fit and the $t_o$ and $t_1$ velocity flow deflection boundaries used in the model fit.  Unfortunately, velocity flow deflections for MC2 are not reliable at STB because the ICME speed in this region at STB is slower than the surrounding solar wind speed. 

We apply the non-force-free flux rope model to the ACE and STB magnetic field data shown in Figures 6 and 7 using the same boundaries for the flux ropes as indicated in Kilpua et al., (2009).  Resulting single spacecraft flux rope inversions at ACE and STB indicate highly inclined flux ropes relative to the ecliptic plane. Details of the single spacecraft fits and the associated can diagrams for this event series are found in Reinard et al., (2010).

\subsection{Synoptic ICME Model Inversion}

As discussed in Section 2, the velocity flow deflection around the ICME obstacle can be used to estimate an impact parameter and normal vector to the leading ICME boundary, assuming a cylindrical shape for the obstacle.  Since the magnetic field observed at STA for MC1 does not have a clearly identifiable flux rope signature, we can employ velocity flow deflections around the ICME to determine its orientation at STA. As a preliminary analysis, looking at the STA velocity components for MC1 (orange region) in the bottom three panels of Figure 7, the flow deflection ahead of$t_o$ is initially westward, indicating STA is on the west side of the ICME symmetry axis, similar to the trajectory T2 shown in Figure 1b. At the leading edge of the ICME, the flow deflection clock angle given by equation (1) gives 19$^{\circ}$, which is generally consistent with the single spacecraft model clock angles at ACE of 35$^{\circ}$ and STB 435$^{\circ}$ (see Reinard et al., 2010). The smaller angle at STA may indicate a gentle distortion of the cross-section of the rope from a high-inclination (19$^{\circ}$) at STA to a less high inclination (43$^{\circ}$) at STB.

The results of the synoptic ICME inversion for MC1 using the magnetic field data from Wind, ACE, and STB along with the obstacle normal constraints from the velocity flow deflections at STA are shown  in the can diagrams Figure 8a and 8b. These are 3-D representations of the flux rope structures within the ICMEs.  Cuts through these structures at each spacecraft correspond to the black dashed lines in the magnetic field time series in Figures 6 and 7. Two different views are shown in Solar-Ecliptic coordinates. The orange can represents the flux rope portion of the MC1 ICME with its size, orientation, and relativity to the Sun clearly indicated. (The orbit of the Earth is shown as the green ellipse.) In Figure 8a, a dipole field line tangent to the axial field in the ecliptic plane threads the rope and serves as a reminder that these structures have magnetic footpoints rooted on the solar surface. In both 8a and 8b, tracks from all four spacecraft through the structure are shown.  The flux rope axis is quasi-perpendicular to the ecliptic plane with a mean clock angle of 39$^{\circ}$ (meaning a 39$^{\circ}$ clockwise rotation from the z-axis in the plane of the sky, as viewed from the Sun). Similar to the single spacecraft results in Reinard et al., (2010), the resulting flux rope structure is highly inclined to the ecliptic plane at ACE, Wind, and STB. (Note the trajectory of STA does not intersect the flux rope.) Unlike the single spacecraft fits, the flux rope is shown to have an elliptical structure, is elongated azimuthally by a ratio of 1.86, and the cross-section of the rope is slightly distorted between STB and ACE through a clock angle of ~8$^{\circ}$. (If we include data from STA containing information about the non-rope portion ICME cross section, the distortion of the cross-section of the entire ICME envelope increases to 24$^{\circ}$.)  

Performing a similar fit for the MC2 ICME, requires use of magnetic field and velocity flow deflection data from ACE, Wind, and STA.  Unfortunately, STB does not exhibit an identifiable flux rope magnetic field signature during this period.  Equally unfortunate is that this MC2 ICME region also has the slowest solar wind speed of the entire 21-24 May 2007 interval. Because of this slow speed, the surrounding solar wind may actually be overtaking the ICME at the STB location, and thus the velocity flow deflections around the obstacle are unreliable. Using the data at Wind, ACE, and STA results in a synoptic model fit shown by the dashed lines in Figure 6 and 7 (for ACE and STA) and represented by the blue-green can in Figure 8a and 8b.  In the case of MC2, the flux rope axis is also quasi-perpendicular to the ecliptic plane, but with a mean axial orientation having a clock angle of 161$^{\circ}$. This orientation indicates the rope's axial magnetic field is pointing southward, and that the two rope axes are nearly perpendicular to one another.  This flux rope is also shown to have an elliptical shape (stretched azimuthally by a ratio of 1.2) and the cross section distorted between ACE and STA by an angle of over 30$^{\circ}$. As shown in Figure 8b, the trajectory of STB does not intersect the rope.

Figure 8c, adapted from Figure 1 of Ki;pua et al., (2009),  plots the Grad-Shafranov reconstructions for MC1 (orange) and MC2 (blue) in the ecliptic plane, with the colored ovals the expected elongation of the ICME boundaries (magnetic field contours from the Grad-Shafranov fits are shown interior to the ICME boundaries).  Comparing Figures 8a-8c, the Grad-Shafranov and synoptic model results are generally consistent, for both MC1 and MC2.  In MC1, STB passes to the east of center, while ACE and Wind pass to the west of center of the flux rope. STA just grazes the edge of the ICME structure, with an impact parameter passing beyond the edge of the flux rope in the ICME interior. Comparing elliptical scale size ratios (major to minor axes) the synoptic result gives 1.84 9 1.200 for MC1 (MC2), consistent with the GSR values of 2.25 (1.15) for MC1 given in Kilpua et al., (2009).

Although the panels in Figure 8 clearly indicate where the magnetic flux ropes, MC1 and MC2, are relative to the spacecraft, it is difficult to picure how the non-flux rope ICME signatures are incorporated into the overall ICME structure  (at STA for MC1 and at STB for MC2).  The elliptical ICME boundaries  shown in Figure 8c were qualitatively suggested by Kilpua et al., (2009) using the observed size of the flux ropes in the radial direction, impact parameters, and axial field directions from the GSR analyses. In the next section, the spatial mapping procedure will remove much of the remaining uncertainty and reveal the impact of the solar wind environment, both inside and around the coherent magnetic flux rope structure. This will enable us to know the posistion of the flux rope relative to the ICME envelope and construct a more quantitative description of the large-scale ICME structure of the event series.

\subsection{Analyses of the 21-24 May 2007 events using spatial maps}

Much of this section focuses on placing the results from the synoptic modeling of MC1 and MC2 in context with the spatial maps of the ambient solar wind. By putting these techniques together we are able to reconstruct an ecliptic plane view of the large-scale ICME geometry for each event and analyze it within the larger framework of the surrounding plasma environment. What is distinctly new about this composite technique is that it reveals aspects of the flux rope and ICME structure and dynamics in a fashion that is far more visually intuitive than individual time-series data tracks alone.

Figure 9 shows the results of these techniques applied in combination to the 2007 May 21-24 time period, including the MC1 and MC2 ICME events.  As described in Section 3, these maps extend a few tenths of an AU upstream and downstream of the two ICMEs, equivalent to observations made from several days earlier to several days after ICME passage. The maps are centered on a point in time that is nearly halfway between MC1 and MC2 as observed at L1. This time (2100 UT on 22 May 2007) is defined as the time ($t_o$) which maps to position R= 0 (x=y=0) on the spatial grid, meaning the origin of the spatial map is located at the Earth's position in the y direction and centered between the two flux ropes in the x direction. Similar to Figure 4, the Sun is off to the left, time increases to the left, and spacecraft tracks through the maps are shown in grey, with the STA and STB tracks bounding the edges of the “visible” region of the ICME structures and solar wind.  In the top panel, and overlaid on the maps in panels 2-5, are black arrow vector field regions showing the GSE Bx and By components of the interpolated magnetic field  (constructed from the fits to the highlighted regions MC1 and MC2 shown in Figure 6 and 7 at STB, ACE, Wind, and STA).  In panels 2-5, the spatial maps for solar wind Np, Vp, Tp, and the He$^{+2}$/H$^+$ ratio are shown for the same interval. Note that for the He$^{+2}$/H$^+$ ratio map, only three spacecraft data are available for interpolation with a much reduced data resolution of 1-hour, resulting in a coarser profile.

\subsubsection{Magnetic Field Analysis}
In order to maintain consistency with Figure 8c, adapted from the Kilpua et al., (2009) study, the top panel in Figure 9 shows the magnetic field interpolation of the two-ICME event series, with the MC1 (MC2) region shaded by the orange (blue) ellipse. The field in thes shaded regions reveal coherent, large-scale rotations consistent with the existence of  flux ropes nested within  larger ICME structures. Note that no attempt has been made to ensure the interpolation of the magnetic field results in a divergenceless field and, as such, the arrows are not intended to represent actual field lines, but rather indicate the location and general direction of the ICME magnetic field in context with the surrounding environment. The extension of the shaded boundaries beyond the STA and STB spacecraft tracks illustrates how far the flux ropes and the surrounding ICME envelopes extend beyond the range of our viewing window. 

In the case of the ICME containing MC1, velocity flow deflections from the synoptic model fit indicate the western edge of the ICME envelope is very close to the STA location (IP$\geq$0.9). Thus it is expected that the western flank of the ICME envelope does not extend much beyond the location of STA, the range of our view. Looking closely at the magnetic field direction at the leading westward flank of MC1, the $B_y$ component, which near ACE is pointing westward, abruptly reverses direction approximatley halfway between ACE and STA.  The location of this reversal in $B_y$, at which the field begins to point in the opposite sense as the field within MC1, delineates geometrically for the first time, the magnetic boundary of a flux rope region within a larger ICME envelope.  In this instance, the MC1 flux rope is revealed as a sub-structure of a larger ICME envelope.  The eastern edge of this envelope is not visible because the edge of our viewing window extends only down to STB, which traverses barely east of center of the MC1 flux rope. 

Turning our attention to the magnetic field of MC2, the spatial map in the top panel of Figure 9 shows the westernmost boundary of MC2 extends beyond our viewing window.  In the synoptic model fit, there is consistency between the impact parameter given by the velocity flow deflections at STA and the impact parameter given by the magnetic field, indicating the large-scale ICME boundary probably does not extend much beyond the flux rope itself. Unfortunately, for MC2 we do not benefit from having velocity flow deflections at the location of STB, where the magnetic field signature is non-flux rope like so we cannot project how much beyond the eastern edge of the viewing window the ICME containing MC2 extends. However, this may prove unnecessary as there are other indicators within the ambient solar wind plasma that give us clues as to where the ICME envelope for MC2 extends in this case. Though our viewing window is constrained to only a small portion of the overall structure of these ICMEs, it is still possible to discern much information about the large-scale geometry and dynamics of these events by looking at the plasma environment during this period.

\subsubsection{Solar Wind Plasma Analysis of the MC1 ICME}
Panels 2-5 of Figure 9 allow us to concentrate on the structure of the plasma environment within and surrounding the ICMEs. Concentrating first on the unusual distortion of MC1in the radial direction, the proton speed in panel 2 reveals a higher-speed region near the ICME trailing edge at STA compared to the other spacecraft. That this portion of the ICME is embedded in a region of higher speed plasma may explain why the ICME is first observed at STA and may be why the western flank of this ICME leads its apex, looking very unlike typical kinematic flux rope model cross-sections (e.g. Owens et al. 2006).  This increased speed also compresses the tail end of the ICME at STA, perhaps contributing to the difficultly of observing anything resembling a flux rope structure at this location. Along with the grazing incidence, the compression of the ICME due to the increased solar wind speed may also explain why the observation at STA is so much shorter in duration than at ACE or STB.  The third panel in Figure 9 shows the proton density is depressed near the core of the MC1 flux rope, with an extended enhanced density region trailing the flux rope proper. $N_p$ is strongly enhanced at the compression region at the trailing edge of the ICME at STA, but the fine-scale structure in this trailing region is poorly resolved. In fact, returning to Figure 7, much of the fine-scale structure at STA during this period (approximately 1900 UT on 22 May 2007 to 1200 UT on 23 May 2007) has been averaged over by the interpolation process. The region shows enhancement, but the details are lost. (To the degree the interpolation fails to capture the correct density profile in this region is discussed in Section 6.) Still, the qualitative detail is enough to confirm that the bulk of the dense material is near the trailing end of the ICME.  What is also interesting is that $N_p$ (and to a lesser extent the He$^{+2}$/He$^+$ ratio in panel 5) is also elevated in the ambient plasma at the trailing edge of MC1 as observed by ACE, Wind, and STB.  This elevated $N_p$ may be indicative of the filamentary material often observed at the back end of CMEs and at the trailing boundary of ICMEs. Taken together, the spatial maps are indicating that a significant portion of this material is external to the associated flux rope structure, rather than co-located with the trailing edge of the rope. This suggests the trailing portion of the ICME envelope extends well beyond the coherent flux rope sub-structure MC1.

As is typical, the proton temperature, $T_p$ in panel 4, shows the coldest temperatures inside MC1, with elevated temperatures in between the two ICME events. There is also a hint of elevated $T_p$ at the trailing edge of the MC1 ICME at STA, consistent with a compression region at this location, but with a much smoother overall profile than the density. 

Moving to the spatial map containing the He$^{+2}$/He$^+$  ratio in panel 5,  we see some interesting features. Typically, the helium density is elevated in the ICME sheath and envelope regions and is generally depressed in the flux rope interior. However, near the leading edge of MC1 at STB, the He$^{+2}$/H$^+$ ratio is elevated over typical solar wind and these levels remain high even as STB nears the core of the flux rope. This enhancement is also observed at ACE in the latter portion of the rope. Another interesting feature at ACE is that the helium enhancement at the trailing edge of MC1 continues, unbroken into the second ICME containing MC2.  This interconnection, which occurs at both STB and at ACE, may be an artifact of the interpolation process or it may mean that the solar source region never fully``'closed''  after the first CME eruption, resulting in a single, extended envelope connecting both events. CME interaction or merging in the solar wind is another possible explanation for this observation (e.g. Gopalswamy 2001, 2004).  However, above ~20 $R_s$ one could argue that the frozen-in flux condition should result in the two ICME envelopes being distinct from one another with a highly compressed parcel of ambient solar wind separating the two independent events. In addition, CME merging requires the second CME to have a much higher speed than the first, which is not evident in the solar or interplanetary observations (Li et al., 2008; Kilpua et al., 2009).

\subsubsection{Solar Wind Plasma Analysis of the MC2 ICME}
Examining the plasma environment surrounding the ICME containing MC2, it is clear the extent of the flux rope, and possibly the ICME envelope do not reach the location of STB. Instead, $Np$ in panel 3 shows the easternmost flank of MC2 nested within a large density enhancement that extends the entire radial dimension of the ICME. Just upstream of MC2, the magnetic field in the ecliptic plane reverses direction in $B_x$ (from anti-sunward to sunward), seeming to bend around the density structure.  Although this upstream magnetic field region may not necessarily be part of the ICME envelope, the field vectors have been included in the spatial map because the reversing field signature well defines the leading-eastern boundary of the flux rope.  The peak of this density structure in the spatial map at ACE and Wind is marked by the dotted line in the time-series of Figure 6 and is bracketed by the $t_o$ and $t_1$ boundaries (blue highlighted region) in Figure 7 at STB. Another reason this upstream region has been included is to highlight the role it may play in the distortion of the flux rope cross section. As the magnetic field arrows in panel 1 clearly show, the center of MC2 is located near the trajectories of Wind and ACE, but the synoptic model results indicate ACE and Wind pass on the east side, approximately one-third of the way from the rope center. This difference is best illustrated by noting the location of the center of the blue oval in panel 1 is not co-located with the center of the flux rope as indicated by the curling magnetic field. In the magnetic field spatial map it appears as though the eastern half of MC2 has been compressed. That the model indicates the flux rope cross section is distorted through an angle of over 30$^{\circ}$ (see Figure 8) suggests the density enhancement may be compressing the eastern side of the flux rope, perhaps bending its shape out of the ecliptic plane. 

The proton temperature in panel 4 and the proton speed in panel 2 are also consistent with this picture. The high density region at the eastern flank of the ICME containing MC2 is also a region of enhanced temperature and slowest speed. By comparison, the coolest and fastest regions of this ICME are inside the actual flux rope MC2.  Although the difference between the fastest speed inside MC2 ($\sim$500 km/s) and the slowest speed in the density enhancement ($\sim$450 km/s) is not too remarkable, it may contribute to a slight asymmetric compression of the ICME overall structure, as the faster plasma overtakes the slower dense structure ahead on the ICME's eastern flank.

\section{Discussion}

The power of this innovative approach comes from combining two separate techniques into a single in situ modeling tool. If we return to Figure 8c for comparison with Figure 9, it is obvious that more information can be derived from these spatial maps than from previous reconstruction methods. Combing the synoptic model inversions with the spatial mapping for use alongside time-series analyses provides a much more comprehensive view of a large fraction of the two ICMEs given as examples. The result is a multi-dimensional reconstruction that reveals many long-sought-after characteristics of ICMEs in a visually intuitive manner; this includes resolving flux rope sub-structure within the larger ICME structures, determination of the ICME envelope boundaries, and the distortion of these ICMEs due to interaction with plasma structures in the surrounding solar wind. 

Another benefit of this approach is that it makes multispacecraft identification of ICMEs much easier than by using individual time series alone. For the highly inclined ropes as MC1 and MC2, magnetic flux rope field rotations are extremely easy to locate as are the boundaries of these ICME substructures. Kilpua et al., (2009) noted that if only observations at L1 had been available, MC2 would not have been identified at all and the identification of MC1 would have been difficult. From Figure 9, the coherent rotations in the flux rope magnetic field signature are clearly recognizable and stand-out easily among the ICME envelope magnetic field and the surrounding ambient plasma signatures.  In fact, the spatial mapping of the magnetic field made possible the refinement of the ICME flux rope boundaries indicated by the vertical lines in the time-series of Figures 6 and 7. Indeed, MC2 is easily recognizable at L1 as is MC1, regardless of whether the magnetic field mapped in Figure 9 is interpolated from the synoptic model results or taken straight from the in situ measurements of the magnetic field at each spacecraft.

A consequence of this composite technique is that the reconstructed flux rope cross-sections show a more realistic distortion when compared with results using the GSR technique.  As discussed by Riley et al. (2004), the GSR technique has difficulty capturing the true distortion of simulated flux ropes, presumably because the method assumes that the structure is in approximate magnetostatic equilibrium. Kilpua et al., (2009) confirmed that the GSR technique successfully fits the ropes MC1 and MC2.  However, the solution did not return a cross-section consistent with the distortion observed at the different spacecraft locations.  In contrast, the methods presented in this paper are essentially empirical; there are few limiting assumptions: flux rope topology at one (or more) spacecraft, an ICME geometry that is locally elliptical or cylindrical in shape, and frozen-in-flux convection of the plasma and field. Our method does not assume any equilibrium state. Thus, the cross-sections presented in Figure 9 are consistent with the size, shape, and arrival time indicated by the observations at each spacecraft location. In the case of MC1, the spatial map reveals an asymmetrical distortion causing the rope flank (and thus the edge of the ICME proper) to lead its apex. Geometries such as these are not allowed by current static, quasi-static, or kinematic models of flux ropes.  Only MHD simulations show these kinds of distortions.

One of the most outstanding implications from performing this synoptic ICME analysis on the 21-24 May 2007 ICME events is that the result argues for a non-closing of the solar source between two eruptions.  In the Li et al., (2008) study, the fast CME on 19 May 2007 appeared to be part of a complex eruption with a second, slower CME observed very close in time and space. The authors found these two CMEs difficult to separate in coronagraph measurements so it was unclear if they were part of the same eruption.  Kilpua et al., (2009) concluded that in-situ observations by STB and Wind supported the interpretation of a complex eruption due to an interval of counter-streaming electrons and a small magnetic cloud-like region immediately following the trailing boundary of MC1.  In the bottom panel of Figure 9, the He$^{+2}$/H$^+$ ratio remains enhanced between the two ICMEs. If we take this to mean that the solar wind in between these two structures is more typical of ICME envelope material than solar wind, then this argues for the two ICME structures being embedded within a single, larger envelope, consistent with being component parts of a multi-eruption composite or complex ICME event (Gopalswamy et al., 2001, 2004).

As will be shown in a companion paper (Reinard et al., ApJ submitted 2012), spatial mapping of the Fe charge states confirm the existence of ICME-like plasma (enhanced $Q_{Fe}$) connecting MC1 and MC2 at multiple spacecraft locations. Thus, these maps give us clues as to how the solar source region responds when CMEs erupt into the solar wind. The continued ``leakage'' of elevated helium density (and creation of high Fe charge states) may be indicative of the solar source region remaining magnetically open to the solar wind after the traditional flare signature associated with CME expulsion.  It also suggests the presence of continued eruption- (or post-eruption-) related heating that serves to pre-condition the source region environment prior to the second eruption. In fact, recent numerical simulations of magnetically coupled sympathetic CME eruptions by T\"{o}r\"{o}k et al., (2011) depict a multi-eruption scenario that almost certainly describes the CME origins for these two May ICME events. If one compares the pre-eruption potential field source surface (PFSS) extrapolation in Figure 4 of Li et al,. (2008) to the magnetic topology and evolution of the T\"{o}r\"{o}k et al., (2011) simulations (their Figure 3), one can immediately associate the first eruption above the west neutral line (WNL) and the second eruption above the center neutral line (CNL) with the simulated flux ropes labelled FR2 and FR3, respectively, in the T\"{o}r\"{o}k et al., (2011). In addition, the T\"{o}r\"{o}k et al. figure panels 3(e) and 3(f) show that the flare current sheet formed during the FR2 eruption acts as the ``breakout'' reconnection that facilitates the sympathetic FR3 eruption. Implications of this scenario are that the elemental and ionic composition, usually associated with the dynamic topological opening of low-lying flux and eruptive flare heating could be present, not just in the interior of each of these ICME flux ropes, but between them as well, precisely what the spatial mapping results show here and in the Reinard et al., (ApJ submitted, 2012) analysis.

\section{Future Work}

There are many aspects of this composite modeling technique that have yet to be explored. It is understood that the spatial mapping of low-inclination flux rope orientations relative to the ecliptic plane will be more difficult to interpret since much of the cross-section will be perpendicular to the spatial mapping plane, making interpretation more difficult. Other unknowns include errors in determining the obstacle normal since the accuracy of using flow deflections around an obstacle is not yet well quantified.  Future work will include using coplanarity, mixed mode, and velocity coplanarity computations to help provide an estimate of the errors to the obstacle normal (e.g. Russell et al., 1983; Schwartz, 1998; Paschmann and Sonnerup, 2008). 

As discussed in Section 3.2, propagation of error is another factor that needs to be quantified for the spatial mapping technique, as is interpolation in Cartesian space as opposed to a spherical coordinate system.  From the limited analysis in Section 3.4, these errors are expected to be small compared to the length-scale of the interpolations and the ICME structures under study. However, there are obvious limitations to the mapping as exemplified in panel 3 of Figure 9. Although enhancements in $N_p$ are observed at all four spacecraft at the trailing edge of MC1, the mapping algorithm interprets these as having local maxima at each spacecraft location, making the overall density enhancement appear "clumpy." This type of artifact results from the algorithm constraining the length-scale of interpolated features in the transverse dimension to be on the same order as the length-scale of the (observed) features in the radial dimension. Thus, instead of a single density enhancement with a much larger transverse dimension as compared to radial dimension (which is probably more representative of reality), the enhancement in Figure 9, panel 3, appears "clumpy" as if several independent enhancements occur at each spacecraft.  Future work will include the quantification of such local artifacts and a modification of the algorithm to use an analytic description of flux rope magnetic fields (obeying $\nabla \cdot B = 0$ at all locations within the grid). Testing the velocity flow deflections with MHD simulations will also help quantify flow deflection errors in obstacle normal determinations.

\section{Conclusions}

In this paper we have developed two new techniques that greatly aid in the modeling and analysis of multispacecraft in situ observations of ICMEs. The first of these techniques improves upon the Mulligan and Russell flux rope model by incorporating velocity flow deflections around the obstacle that provide a measurement of the large-scale ICME orientation.  This additional information relaxes the constraint requiring the model to fit only flux rope structures and opens the possibility of modeling a larger class of ICMEs. The second technique involves creating spatial maps of multispacecraft data by intelligently interpolating the data in the intervening space between the known data tracks. Combining these two techniques together, we have created a powerful composite modeling tool that complements and greatly extends the utility of more traditional time-series analyses. Besides successfully reconstructing the ICME structure previously reported with existing models, it is now possible to go beyond these models to quantitatively reveal substructures, spatial distortion, and dynamics in the large-scale ICME structure, and place them in context with the ambient plasma environment. What is also distinctly new about the approach in this paper is that it reveals aspects of flux rope and ICME geometry and dynamics in a way that is far more visually intuitive than individual time-series data tracks alone. Although the assessment of numerical errors associated with this new model have only begun to be probed, studies are already underway to better understand the limitations using different spatial and temporal resolution data and MHD simulations. The results of this study will be included in a future publication.

Consequences of ICMEs are far-reaching in the heliosphere and the ability to exploit multi-spacecraft plasma and magnetic field data as demonstrated in this paper creates a deeper understanding of these structures in context with the plasma environment in which they are surrounded. Utilizing data in this way gives us clues to how CMEs are formed and ejected into the heliosphere, how ICMEs are structured and evolve in the solar wind, and what is the relationship between solar and in situ observations of these phenomena.  Overall, it is this better understanding of the CME-ICME connection that is of key importance for solar-terrestrial research and space weather forecasting.




%




%




%
\appendix
\section{Delaunay Triangulation and Voronoi Diagrams}

In general, triangulation is a type of discretization of a set of points defining, say a geometric object, into  a number of smaller spatial elements called simplices. These simplices contain a minimum number of vertices $(d+1)$ for the dimension in which they are defined $\mathbb{R}^2$. In two dimensions, a triangulation of a set of points $P \subset \mathbb{R}^2$ can be thought of as a function $T$ that operates on $P$ and returns a surface comprised entirely of triangles $T(P)$.  The convex hull of $P$ is defined as the convex polygon that traces the perimeter of the points. Figure A1 shows an example of a convex hull in two dimensions. The concept is most easily illustrated by imagining an elastic band being placed around the collection of points and having it constrict until it wraps tightly around the outermost points in the set.  The area bounded by the band is the convex hull.

In a triangulation, the convex hull of $P$ is subdivided into simplices such that any two simplices intersect in a common face or not at all. The set of points used as vertices of the subdividing simplices is $Q \subseteq P$. Returning to the two-dimensional case, for each triangle on the surface, a unique circle can be defined such that it passes through all the vertices of the triangle. These circles are known as ``circumcircles'' (because they circumscribe the vertices of their respective triangles). If the circumcircle of a triangle contains no points in its interior, irrespective of how many vertices it has on its circumference, the triangle is called ``Delaunay.'' This means that there can be any number of vertices on the circumcircle, but the circle itself encloses no vertex-- it is empty. Not surprisingly, Delaunay triangulation $DT(P)$ is one that generates a surface comprised only of Delaunay triangles. Such a triangulation for $P \subset \mathbb{R}^2$ is shown in Figure A2. Note that the union of all simplices in the triangulation is the convex hull of the points.

The geometric dual of the Delaunay triangulation is the Voronoi tessellation. Defined mathematically, for a $T$ nonempty subset of $S$, the Voronoi face $V(T)$ is the set of points in $\mathbb{R}^d$ equidistant from all members of $T$ and closer to any member of $T$ than to any member of $S$ that is not in $T$.  This means that the Voronoi face is always a nonempty, open, convex, full-dimensional subset of $\mathbb{R}^d$. In two dimensions, the Voronoi face is a polygon (see Figure A3).  The Voronoi diagram of $S$ is the collection of all nonempty Voronoi faces $V(T)$ for $T \subseteq S$. The Voronoi diagram forms a cell complex that partitions the convex hull of $S$ in a similar fashion as Delaunay triangulation. In fact, these two structures can be easily constructed from one another. The vertices in a Voronoi tessellation are centers of the circumcircles of the Delaunay triangles. Figure A3 shows that by connecting the centers of the circumcircles shown as red dots in the left panel, the Voronoi diagram (right) can be produced. The resulting Voronoi diagram is an extremely powerful discretization tool and is used in many interpolation techniques.  As will be discussed in the next section, Voronoi polygons are useful for determining weights in the natural neighbor interpolation method.

\subsection{Natural Neighbor Interpolation Method}

The natural neighbors of a point x are defined as the neighbors of x in the Delaunay triangulation of $P \bigcup {x}$. Equivalently, the natural neighbors are the points of $P$ whose Voronoi cells would be partially or wholly removed upon the insertion of $x$ into the set. More precisely, let $V_{p_i}$ be the Voronoi cell of $p_i$ in the Voronoi diagram of $P$ and let $Vx$ be the Voronoi cell of $x$ in the Voronoi diagram of $P \bigcup$ {x}. The natural region $NR_{ x,p_i}$ is the portion of $V_{p_i}$ stolen by x, (i.e.~$V_x \bigcap V_{p_i}$.) 
Let $w_{p_i}(x)$ define the n-dimensional volume of $NR_{x,p_i}$ (e.g. an area in $\mathbb{R}^2$ and a volume in $\mathbb{R}^3$).  Then $ NR_{x,p_i}=\emptyset $ and $w_{p_i} = 0$ if $p_i$ is not a natural neighbor of $x$. The natural coordinate associated with $p_i$ is defined by the fractional area:
\begin{equation}
\lambda_{p_i}(x) = \frac{w_{p_i}(x)}{\sum\limits_i w_{p_i}(x)}.
\end{equation}
These natural coordinates have three important properties:
\begin{enumerate}
\item $\lambda_{p_i}(x)$ is a continuous function of $x$, and is continuously differentiable, except at the data sites. 
\item the $\lambda_{p_i}(x)$ are bounded if $x$ belongs to the convex hull of $P$. In two dimensions, this means $\lambda_{p_i}(x)$ vanishes outside the union of the circumcircles circumscribing the Delaunay triangles incident to $p_i$.
\item The $\lambda_{p_i}(x)$ satisfy the local coordinate property (LCP) identity, which states that $x$ is a convex combination of its neighbors:
	\begin{equation}
	\sum\limits_{i}  \lambda_{p_i}(x) p_i = x.
	\end{equation}
\end{enumerate} 

\noindent When $x$ lies outside the convex hull of $P$, $w_{p_i}(x)$ is unbounded if $p_i$ is a vertex of the convex hull. In order to keep the $w_{p_i}(x)$ bounded, the domain over which the natural coordinates are to be computed must be bounded. This can be accomplished by limiting the insertion of point $x$ to a bounding box within the convex hull.
Now assume that each $p_i$ is defined along a continuously differentiable function $h_{p_i}$ in $\mathbb{R}^d$ satisfying $h_{p_i}(p_i)=0$. The natural neighbor interpolation of the $h_{p_i}$ in two dimensions is defined as:
\begin{equation}
H(x) = \sum\limits_i^n \lambda_{p_i}(x) h_{p_i}(x).
\end{equation}
As indicated above, a bounding box $B$ is needed to bound the natural coordinates of any point $x$ to lie within the convex hull of $P$. The set of data points consists of the $p_i$ plus some points $q_i$ added on $B$. Let $h_{q_i}=0$ for all points $q_i$ on the bounding box. Once the interpolation is complete, $P$ will denote the union of the sample points $p_i$ and the $q_i$. For a given $x$, $H(x)$ is easily evaluated once the $\lambda_{p_i}(x)$ have been calculated. 
In the simplest terms, $H(x)$  is the estimate of the function value $h_{p_i}$ at the point $x$, arrived at by summing the function value at the known $p_i$ times the weights $\lambda_{p_i}$, which have been predetermined by the area stolen from the change in the Voronoi faces when the new point $x$ has been inserted into the mix. 

Figure A4 illustrates this process.  The green circles, which represent the interpolating weights in each of the cells ($\lambda_{p_i}$),  are generated using the ratio of the blue shaded area in each cell to that of the cell area of the surrounding points. The blue shaded area represents the new Voronoi cell created after inserting point $x$ into the set. The area comprising this new cell is stolen from existing cells by inserting the point $x$ and recalculating the Delaunay triangles (and thus the Voronoi cells) in the affected area. 

Although the nearest neighbor method is not the only way to estimate the gradients, it has advantages over simpler methods of interpolation in that it provides a continuous interpolation of the underlying function except at the site of the data points ($C^1$ continuous). There exist several other methods developed recently that may provide superior results, such as the Farin’s $C^1$, and Hiyoshi’s $C^2$ methods, which are proving robust under irregular triangulations such as time series data. Such methods may be incorporated in future versions of the model.

\subsection{Application to Time Series Observations}

Figure A5 shows a distribution of points along four simulated spacecraft tracks transformed onto a spatial domain using the technique outlined in section 3.1 using a simulated solar wind speed. Performing a Delaunay triangulation of this spatial data results in the blue triangles connecting the data points along and between the spacecraft tracks.  Triangles within the convex hull (between the two outermost spacecraft tracks) are constructed from points that are in close proximity and sequential in time. Natural neighbor interpolation is expected to work well in this region. Triangles beyond the convex hull (outside the two outermost spacecraft tracks) connect points remote from each other and non-sequential in time. In these regions there is insufficient sampling to accurately capture the surface; thus interpolations in these regions will be spurious and should be ignored. The estimators, $H(x)$, determined from the natural neighbor method are applied to the bulk solar wind parameters $N_p$, $|V_p|$, $T_p$ and $\bf{B}$, corresponding to each spacecraft track on the spatial grid. This results in scalar and vector fields interpolated over the entire spatial domain determined by the multi-spacecraft data points within a bounding box $B$. Because the Delaunay triangulation and the Voronoi diagram are based on the spacecraft observational geometry and relative trajectories, the spatial grid and the the natural neighbor coordinates, $\lambda_{p_i}$,  need to be calculated only once per event. The advantage of this is the continuous functions determined from the multi-spacecraft plasma parameters $N_p$, $|V_p|$, $T_p$ and $\bf{B}$ representing the $h_{p_i}(x)$ can be changed independently, without the need for a re-triangulation. Thus, the solar wind $N_p$,$ |V_p|$, $T_p$, and $\bf{B}$ scalar and vector fields for a single event are all calculated using the same underlying Delaunay triangulation and weighting functions.

\subsubsection{March 25-27, 2007 solar wind contour maps}

To illustrate the usefulness of this technique we perform a Delaunay triangulation and Sibson interpolation of ACE, Wind, STA, and STB plasma data during a moderately disturbed period centered on March 27, 2007, when the STA and STB spacecraft separation was ~3.6 degrees, nearly half their separation distance during the May 22, 2007 event analyzed in Section 4.  Contour maps resulting from the spatial mapping and interpolation processes are shown in Figure A6 for the solar wind parameters $N_p$ and $T_p$. (Contour maps for the proton bulk speed $|Vp|$ is shown in Figure 4.) 

Four separate Delaunay triangulations and natural neighbor interpolations are performed between the various spacecraft data. In the top panel we use data from all four spacecraft (STA, ACE, Wind, and STB) against which we compare the other less constrained maps. The middle two panels use data from three spacecraft (STA, Wind, and STB followed by STA, ACE, and STB).  The bottom panel uses data from only STA and STB in the mapping and interpolation processes. In each panel the Sun is to the left and the units are in AU in the ecliptic plane in Solar Ecliptic coordinates. Grey lines in each panel show which spacecraft data is used and how the spacecraft trajectories track through the spatial domain. As mentioned previously, these contour maps are not a single snapshot in time (i.e. the spacecraft trajectory in the spatial domain takes a finite amount of time to traverse the grid) resulting in features further to the left (nearest the Sun) being later in time than features to the right of the map. Despite this limitation, it is clear that the three- and two-spacecraft interpolations suffer only slight degradation from the four-spacecraft interpolation in the top panel.

By extracting interpolated data in the three- and two-spacecraft contour maps that correspond to the missing ACE and Wind spacecraft tracks (missing grey lines in panels 2-4 in Figures A6 and A7), it is possible to compare the results of the three- and two-spacecraft interpolations to the actual time series data. This is easily accomplished through a reverse transformation of the interpolated data back to the time domain using the same bulk plasma velocity components in the ecliptic plane.

Figure A8 shows the results of three sets of time-series reconstructions for Wind and ACE (the other three sets are shown in Figure 5).  The purpose for this exercise is to determine to what degree the mapping and interpolation processes can recover spatial information between spacecraft data tracks, how well the time-series data can be reconstructed, and quantify the error.  Figure A8 shows three examples of reconstructed solar wind $|Vp|$, $ Np$, $Tp$ data for Wind and ACE. The top panel shows the comparison of Wind data for the contour maps of Vp shown in Figure 4. In the plot, the blue line shows the observed Wind data, while the red line and the black lines show the reconstructed or “modeled” data from the three-spacecraft contour map and two-spacecraft contour map in Figure 4b and 4d, respectively.  Note the similarity of the interpolated data to the actual observations. 

The second and third panels show similar results for the reconstructed proton density and temperature at ACE. The contour maps associated with these two- and three-spacecraft time series model results are shown in Figures A5 and A6.  Note that the error is largest for the more highly variable Np and Tp and lowest for the smoother profile of $|Vp|$. Note also that timing errors and amplitude errors increase as the data moves away from the center time t. This is expected due to the propagation of error in the estimation of spatial grid (see Figure 3) as the mapping moves further from the center time t, which causes increased error, not only in the spatial mapping of the spacecraft locations relative to one another, but also in the estimation of the gradients used for interpolation. 

Overall, the success of the mapping and interpolation processes to reconstruct nearly quantitatively accurate time-series in a moderately disturbed period similar to that encountered during the May 22, 2007 ICME, lends confidence that the method is robust enough to apply to the May 22, 2007 event period. During the ICME event, the spacecraft separation distances between the Earth positioned spacecraft (ACE and Wind) and the STEREO spacecraft will be on the same order as the separation between STA and STB in the March 27, 2007 exercise. This successful demonstration indicates that we can make (at the very least) qualitative assessments of the bulk plasma parameters when interpreting the spatial maps on similar length-scales.


%


%






%








%


\begin{acknowledgments}

T.M. and A.A.R. acknowledge support from NASA SR\&T NNX08AH54G. B.J.L.acknowledges support of AFOSR YIP FA9550-11-1-0048 and NASA HTP NNX11AJ65G. Support for the STEREO mission in-situ data processing and analysis was provided through NASA contracts to the IMPACT (NAS5-03131) and PLASTIC (NAS5-00132) teams. The authors thank the ACE MAG, SWEPAM, and SWICS teams for making their data available on the ACE Science Center Web site (http://www.srl.caltech.edu/ACE/ASC/). T.M. also acknowledges R.A. Leske for his invaluable discussions during the early stages of this manuscript.

\end{acknowledgments}

\end{article}




%


%





%





\begin{figure}
\includegraphics[width=100mm,bb=0 0 1800 1900]{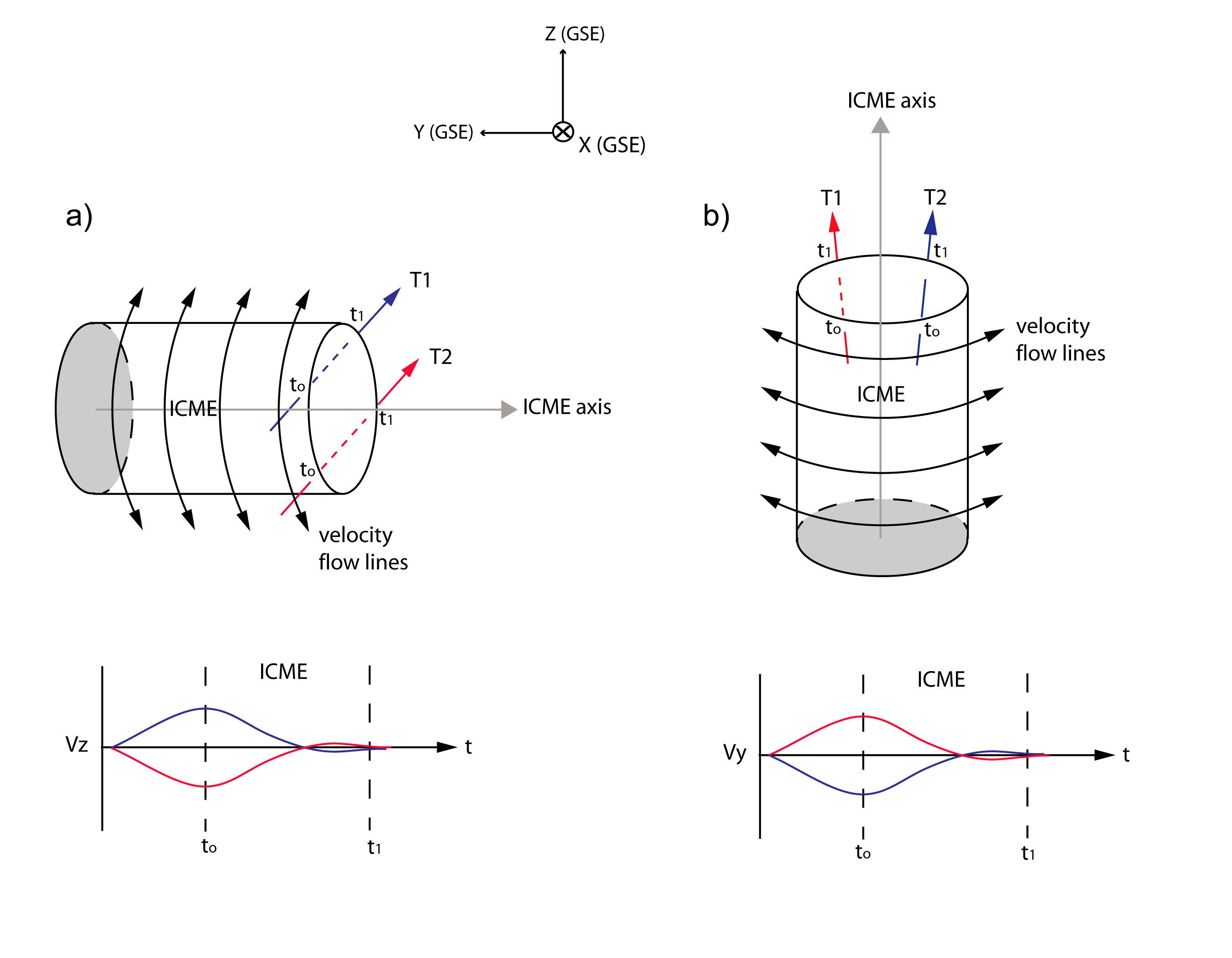}
\caption{Solar wind velocity flow deflections in GSE coordinates for (a) low inclination and (b) high inclination ICME obstacles (adapted from Lindsay (1996).)  For an ICME in the ecliptic plane (left), the flow will be deflected either northward ($+v_z$  for trajectory T1) or southward ($-v_z$ for trajectory T2) of the obstacle as it passes through the upstream solar wind. ICMEs perpendicular to the ecliptic plane (right) will produce only eastward ($+v_y$) or westward deflections ($-v_y$) ahead of the ICME.}
\end{figure}

\begin{figure}
\includegraphics[width=80mm,bb=0 0 1100 2000]{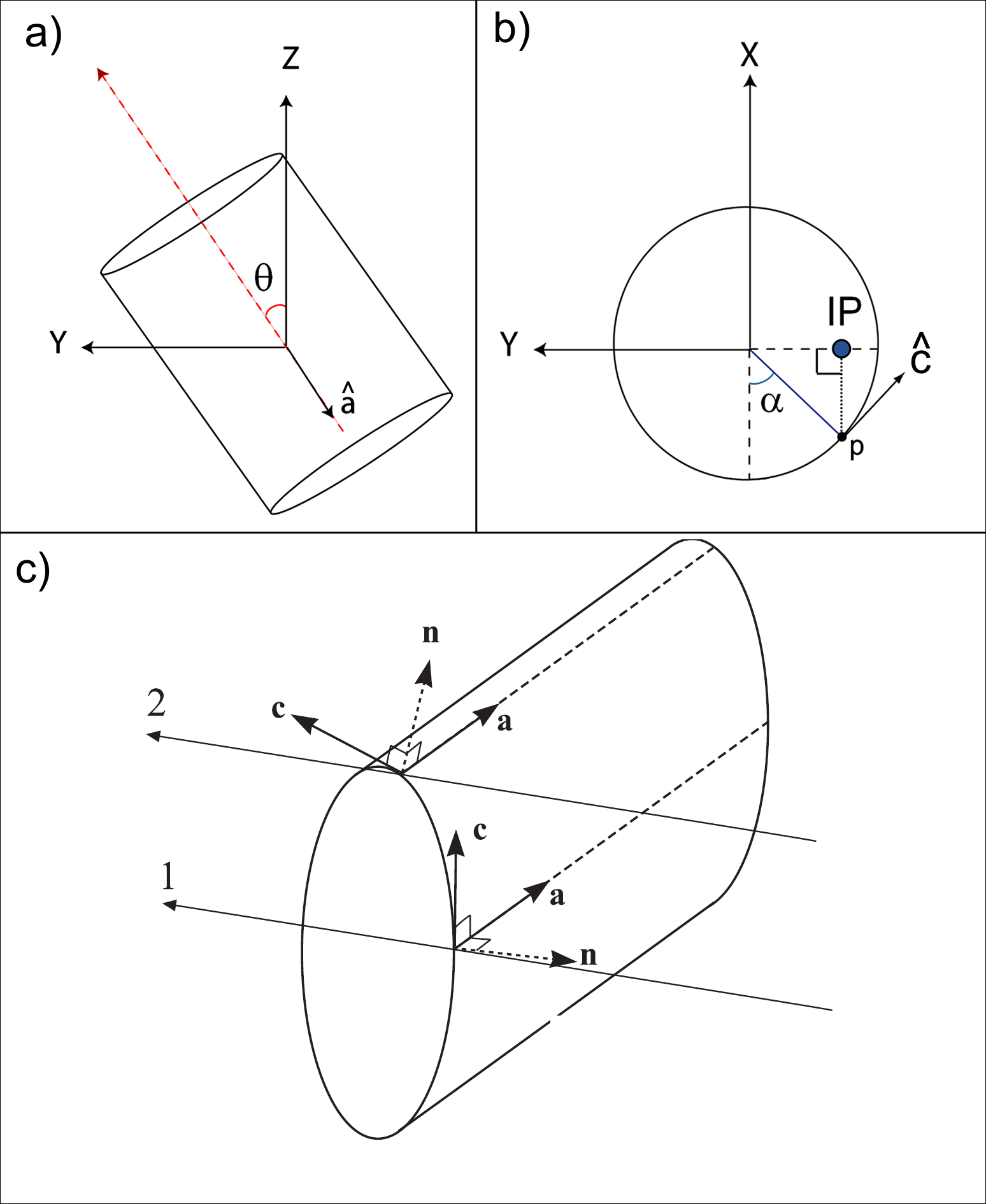}
\caption{Diagrams for the ICME orientation in Figure 1b depicting how velocity flow deflections can be used to determine the impact parameter and ICME obstacle normal using techniques from Lindsay (1996) and Owens and Cargill (2004).  (a) the clock angle  determines the unit vector  $\hat{a}$. (b) the velocity flow deflection is tangent to the ICME obstacle giving vector $\hat{c}$.  (c) the obstacle normal $\hat{n}$ can then be determined from $\hat{a} x \hat{c}$.}
\end{figure}

\begin{figure}
\includegraphics[width=100mm,bb=0 0 200 300]{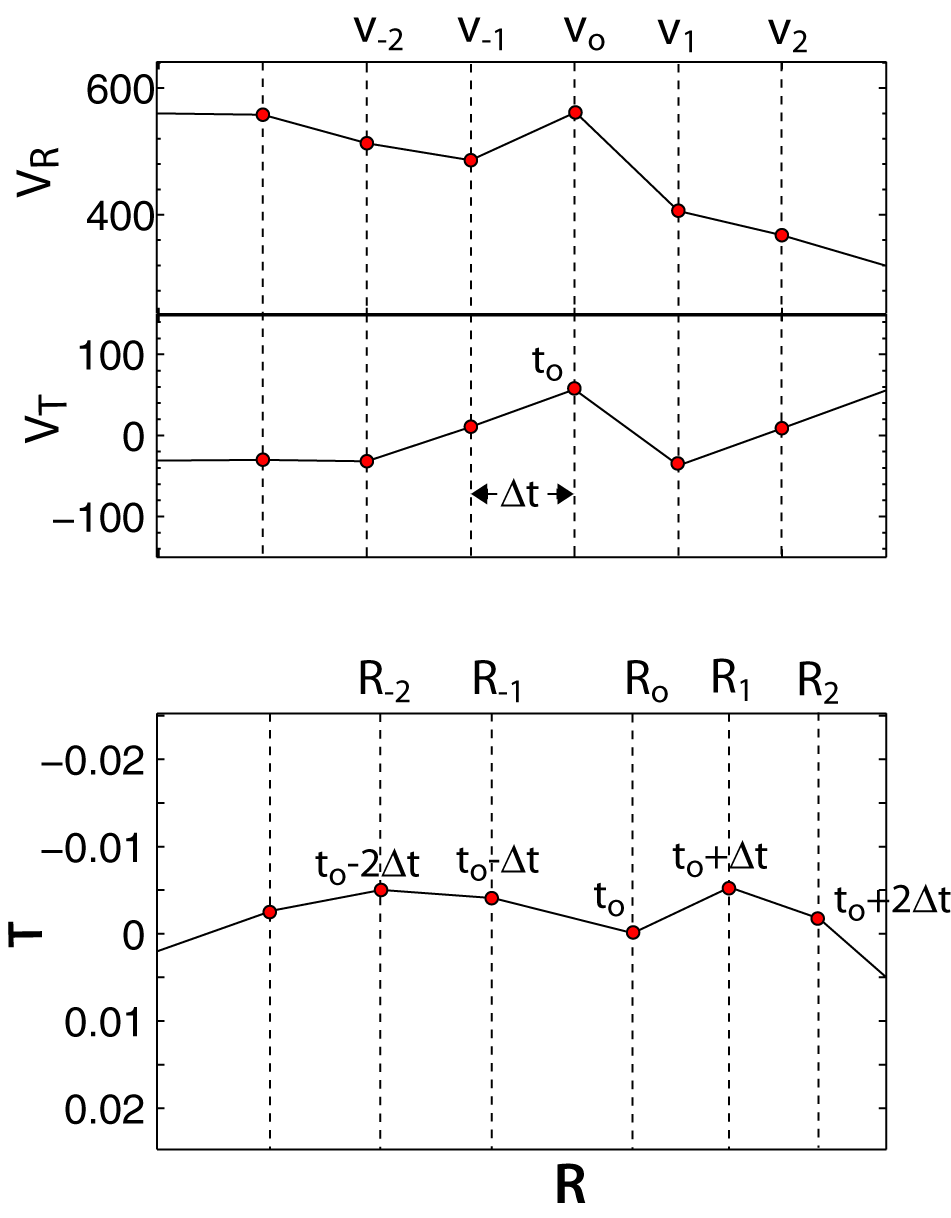}
\caption{Plot showing spatial mapping procedure using time-series velocity component data $v_R$ and $v_T$ (for spacecraft along the Earth-Sun line this is analogous to $v_x$ and $v_y$ in GSE coordinates).  Mapped  data is centered on time t$_o$ to minimize propagation of error from iteration of range calculation. By knowing $\Delta$t between time steps and the velocity components, the spacecraft trajectory at each time step in two-dimensions (R, T) can be determined. Note the sampling frequency of the data affects the resolution of structural features in the data. In this work, 10-minute averaged data is used in order to maximize structural resolution without introducing statistical errors in the Delaunay triangulation due to poorly resolved geometry.}
\end{figure}

\begin{figure}
\includegraphics[width=70mm,bb=0 0 800 2000]{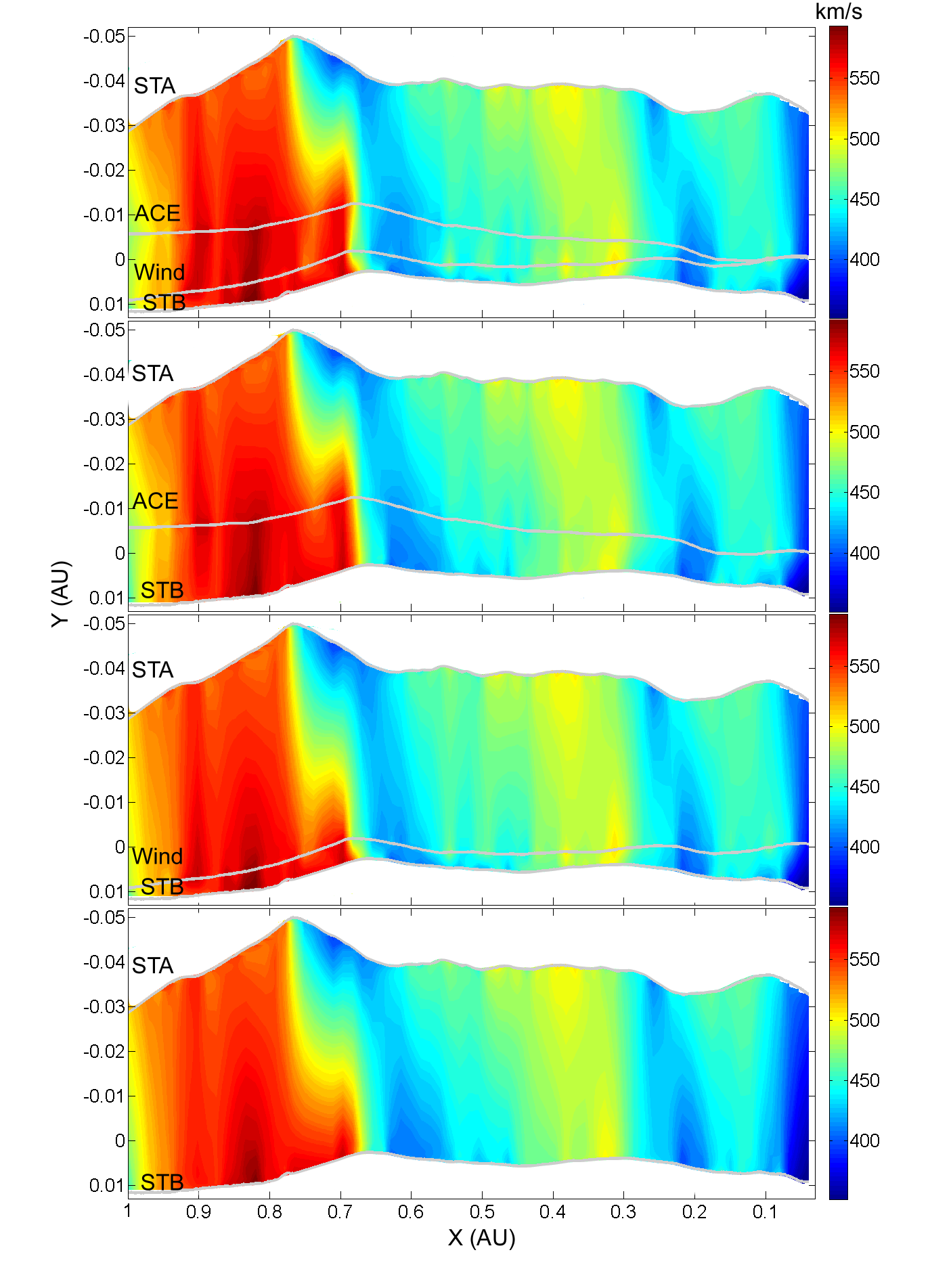}
\caption{Contour maps showing the $V_p$ scalar field interpolations for 24-29 March 2007 using Wind, ACE, STA, and STB data. Top panel is a 4-spacecraft interpolation (STA, ACE, Wind, and STB). The middle two panels use 3-spacecraft (STA, Wind, and STB or STA, ACE, and STB) and the bottom panel uses 2-spacecraft (STA and STB) for the construction of the proton speed contour map. The grey lines show the spacecraft tracks across the map.}
\end{figure}

\begin{figure}
\includegraphics[width=150mm,bb=0 0 800 800]{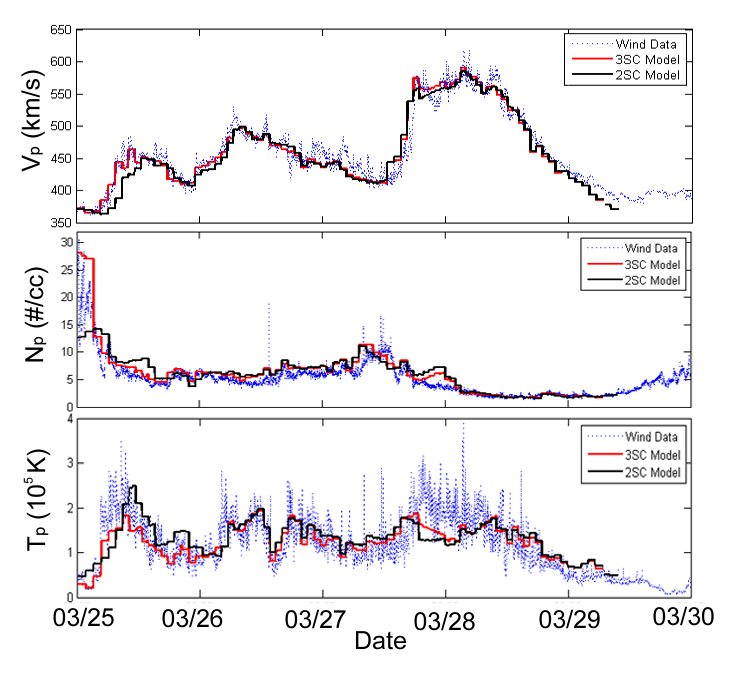}
\caption{Reconstructed time series model tracks using two- and three-spacecraft contour maps created from Wind, ACE, STA, and STB data. Top panel shows reconstruction of Wind $V_p$ data using the contour maps shown in the bottom two panels of Figure 4.  Middle panel shows similar reconstruction for Wind $N_p$ data from the contour maps in the second and last panel of Figure A6. Bottom panel shows Wind $T_p$ data reconstructed from the contour maps in the second and last panel of Figure A7. In all cases the reconstructed model tracks are shown as red and black lines and the original time series data is shown in blue.}
\end{figure}

\begin{figure}
\includegraphics[width=140mm,bb=0 0 800 900,clip]{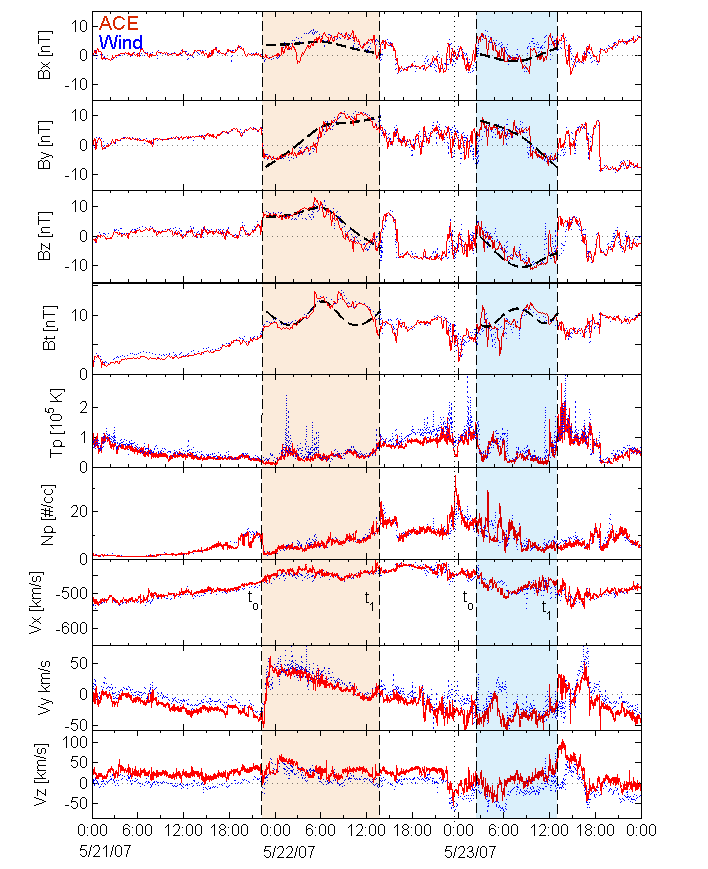}
\caption{Solar wind magnetic field and plasma data for the 22 and 23 May 2007 ICMEs from ACE and Wind. The MC1 (MC2) region is highlighted in orange (blue). The two magnetic flux rope fits at ACE are shown as are the $t_o$ and $t_1$ velocity flow deflection intervals at ACE in a similar fashion as shown in Figure 1b. Flow deflections for ICME corresponding to MC2 at ACE give a clock angle of 190$^{\circ}$. The impact parameter is more difficult to define because the leading ICME boundary has a slower speed than the remaining ICME. In general, the flow deflections for the MC2 ICME are more prone to error due to the slower convection speed for this ICME compared to the ambient plasma. The dotted line at 0 UT on 23 May 2007 coincides with the beginning of the magnetic field vectors for thisICME  plotted on the spatial maps shown in Figure 9.}
\end{figure}

\begin{figure}
\includegraphics[width=150mm,bb=0 0 800 800,clip]{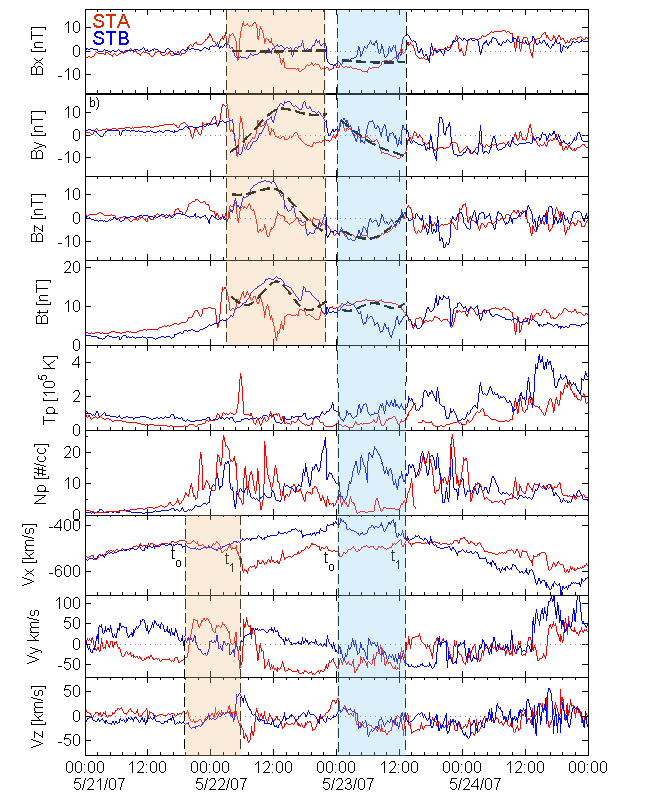}
\caption{Solar wind magnetic field and plasma data for the 22-23 May 2007 ICMEs observed at STA and STB. The MC1 (MC2) region is highlighted in orange (blue). The magnetic flux rope fits at STB (for MC1) and STA (for MC2) are shown,  as are the $t_o$ and $t_1$ velocity flow deflection intervals at STA (for both ICMEs) in a similar fashion as Figure 1b. Velocity flow deflections cannot be used at STB for the ICME containing MC2 because the ICME speed in this region is slower than the ambient plasma.}
\end{figure}

\begin{figure}
\includegraphics[width=180mm,bb=0 0 300 250]{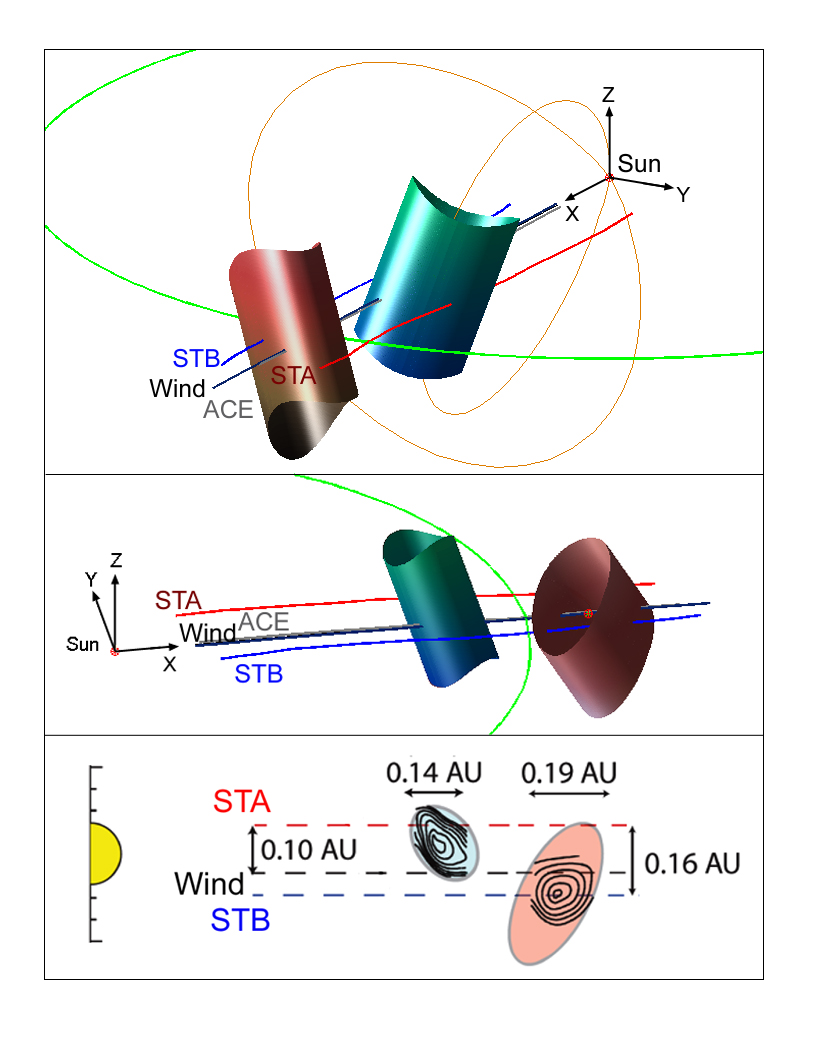}
\caption{Extended flux rope (Mulligan et al., 2011) and GSR results (Kilpua et al., 2009) for the 22 May 2007 flux rope ICME at ACE, Wind, STA, and STB. Top two diagrams show 3D views of the ICME: The stretched cylinders are outer boundaries of the flux ropes on May 22 and May 23 2007. The Earth’s orbit  is  a partial green ellipse. The ropes and spacecraft trajectories are color coded after Kilpua et al., (2009).  Bottom: A rendering of the May 22 ICME (orange) and the May 23 ICME (blue) cross sections in the ecliptic plane adapted from Kilpua et al., (2009).  Note the size, orientation, elongation of the ICME boundaries, and the impact parameters of the spacecraft traversals are similar, but not exactly the same between the two analyses. }
\end{figure}

\begin{figure}
\includegraphics[width=210mm,bb=0 0 800 650,clip]{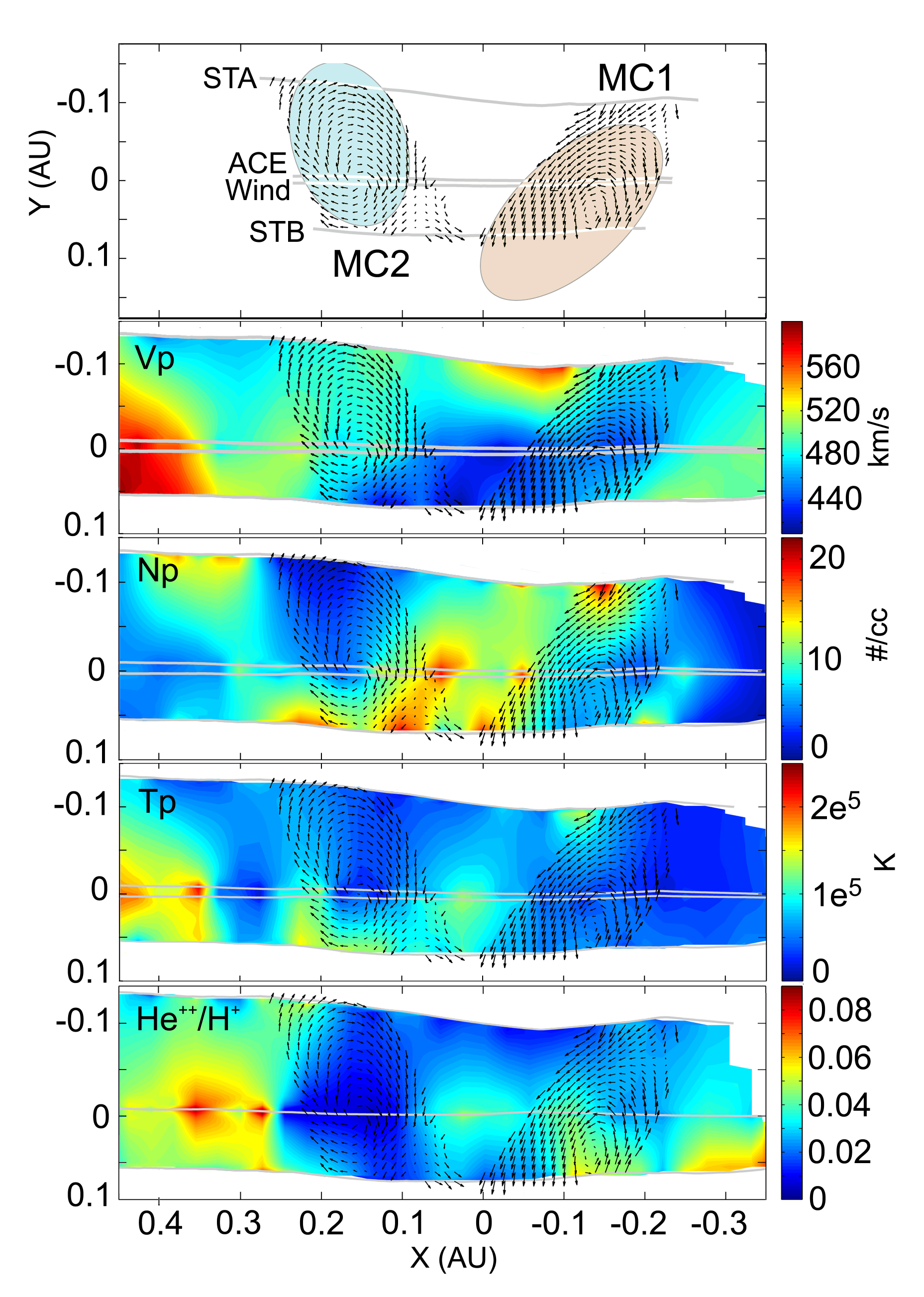}
\caption{Spatial contour maps showing solar wind data interpolations during the 21- 24 May 2007 ICME events. The maps are in the ecliptic plane with the Sun off to the left (in the same sense as Figure 8c). The spacecraft tracks and ICME magnetic field model fits are overlaid on the maps (with flux rope cross section boundaries shown as the blue and orange ovals, respectively). The panels from top to bottom are: the interpolated magnetic field, solar wind $V_p$, $N_p$, $T_p$,  and the He$^{+2}$/H$^+$ ratio within and surrounding the two ICMEs.  }
\end{figure}


\setcounter{figure}{0}
\renewcommand{\thefigure}{A\arabic{figure}}

\begin{figure}
\includegraphics[width=130mm,bb=0 0 1100 1000,clip]{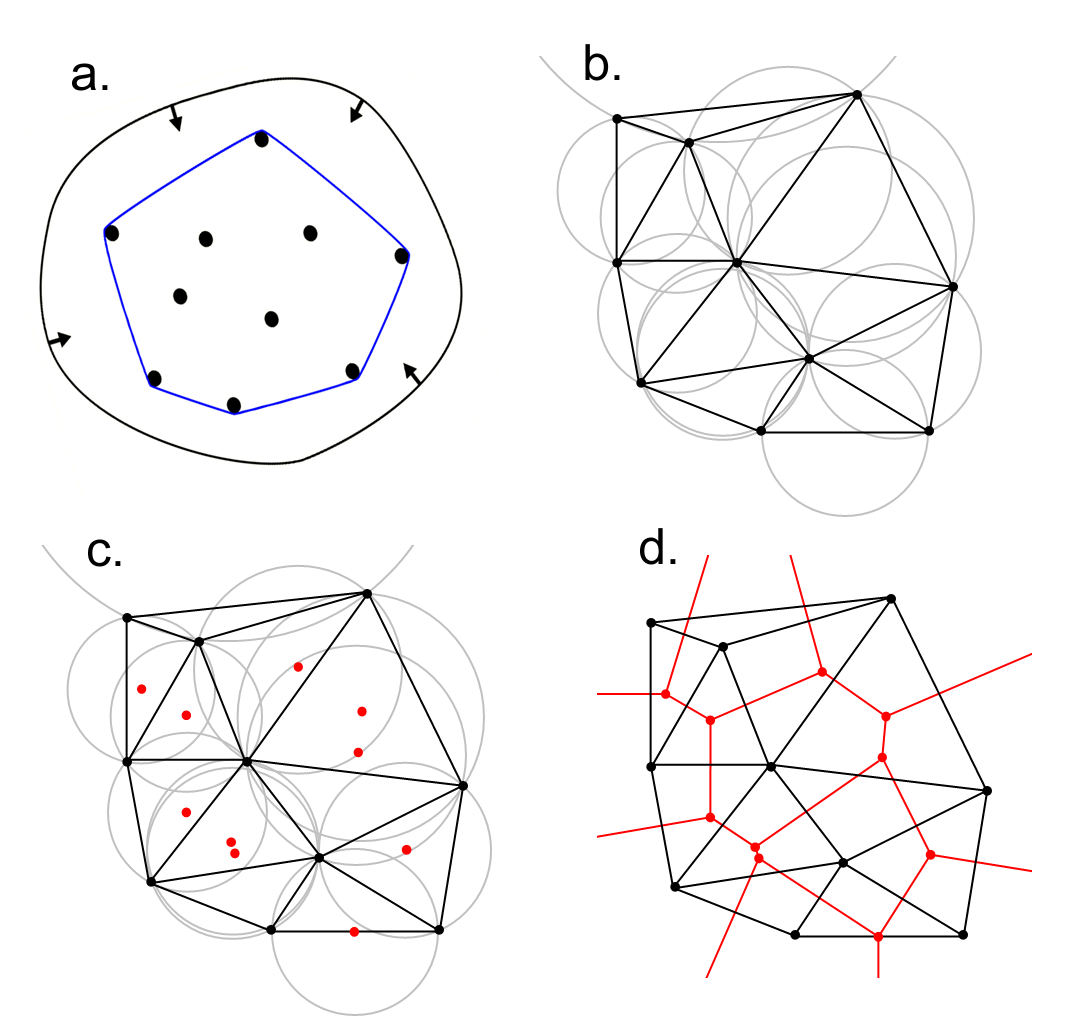}
\caption{(a) shows the convex hull of P. This concept is most easily illustrated by imagining
an elastic band being placed around the collection of points and having it constrict until it wraps
tightly around the outermost points in the set. The area bounded by the band is the convex hull.
 (b) shows a Delaunay triangulation DT(P) is one that generates a surface
comprised only of Delaunay triangles The union of all simplices in
the triangulation is the convex hull of the points. (c) by connecting the centers of the circumcircles shown as red dots, the Voronoi diagram (d) can be produced.}
\end{figure}

\begin{figure}
\includegraphics[width=100mm,bb=0 0 1300 1000]{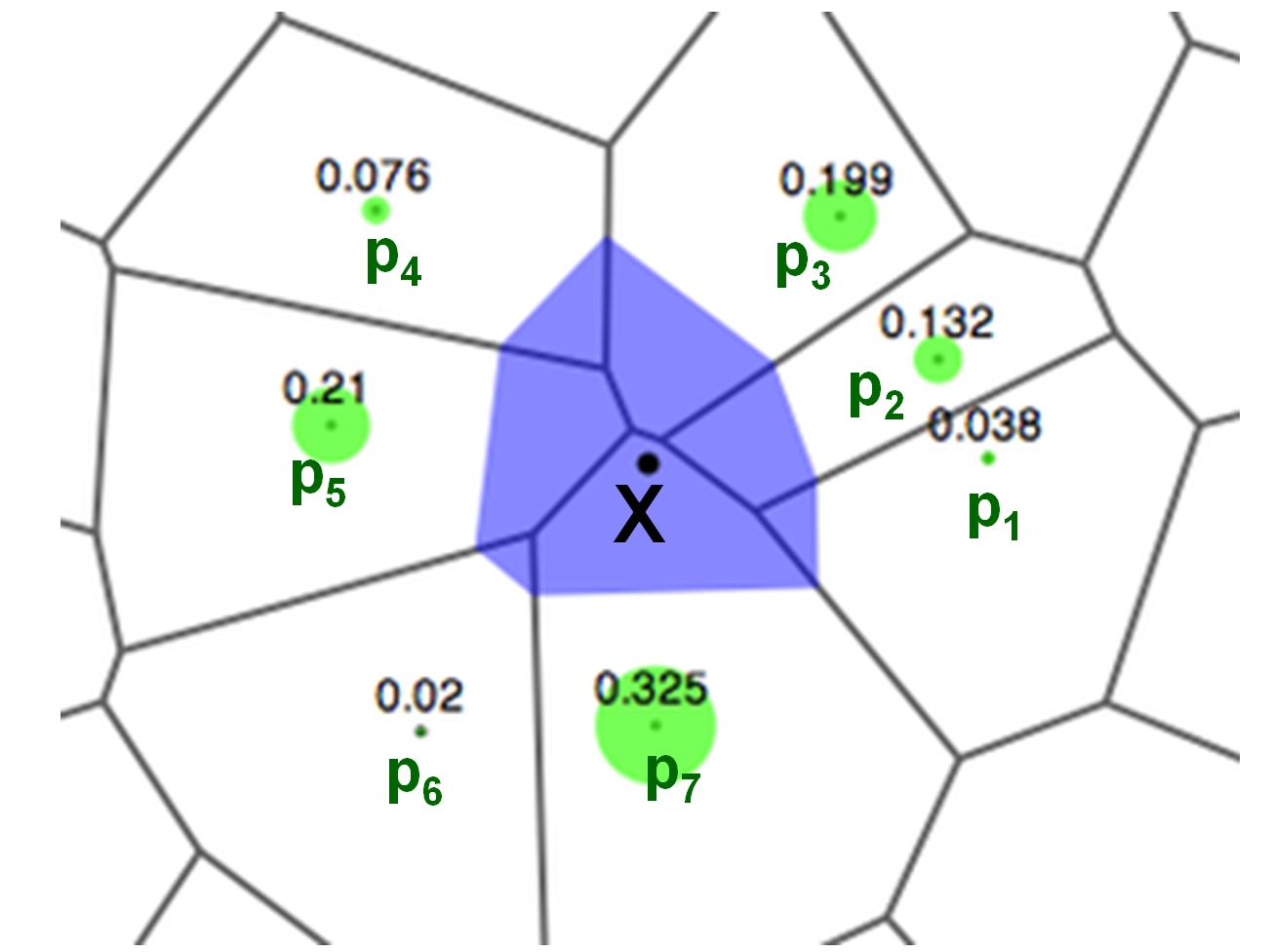}
\caption{ Natural neighbor interpolation. Example of a natural region in 2D. The natural
neighbors of x are $p1,…,p7$. The colored circles represent the interpolating weights, $w_i$, generated
using the ratio of the shaded area to that of the cell area of the surrounding points. The shaded
area is created by the insertion of the point x into the Delaunay triangulation. }
\end{figure}

\begin{figure}
\includegraphics[width=150mm,bb=0 0 900 400]{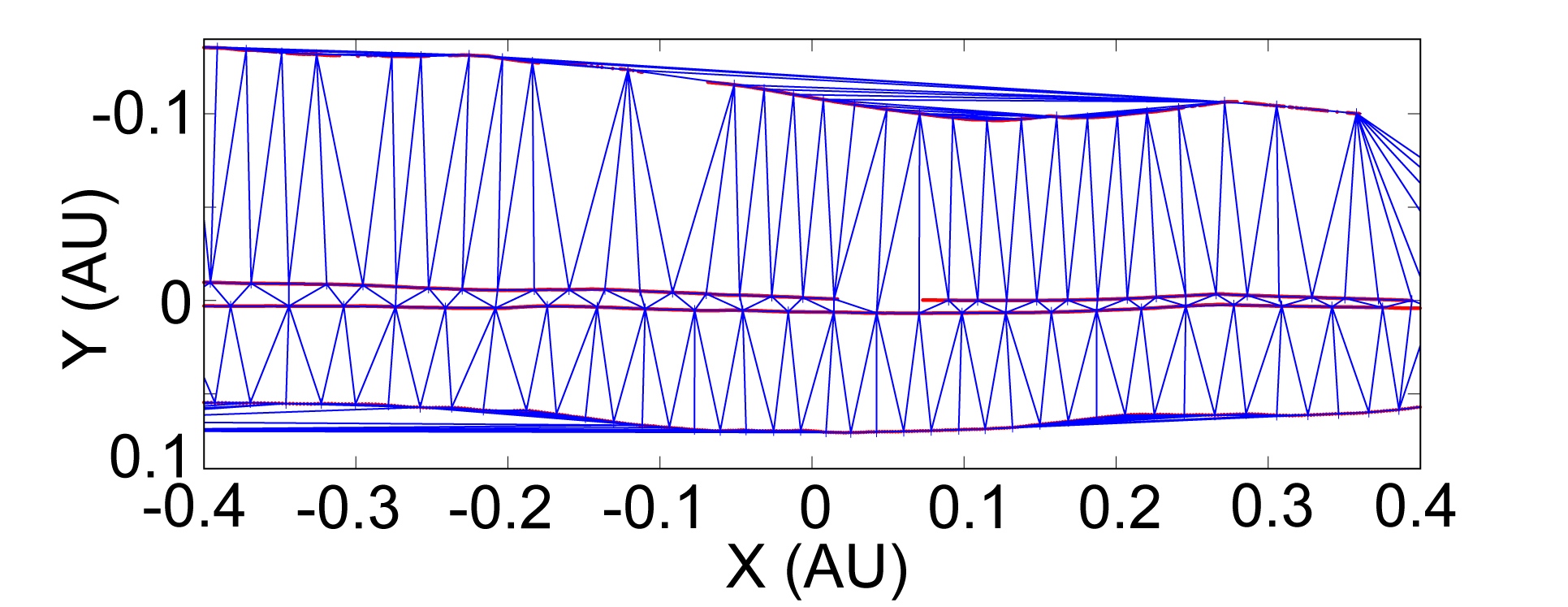}
\caption{shows a distribution of points along four simulated spacecraft tracks transformed onto
a spatial domain using the technique outlined in section 3.1 using a simulated solar wind speed.
A Delaunay triangulation of this set is shown by the blue triangles connecting data points along
and between the red spacecraft tracks. Triangles drawn exterior to the convex hull of the set are
discarded. }
\end{figure}

\begin{figure}
\includegraphics[width=200mm,bb=0 0 450 350]{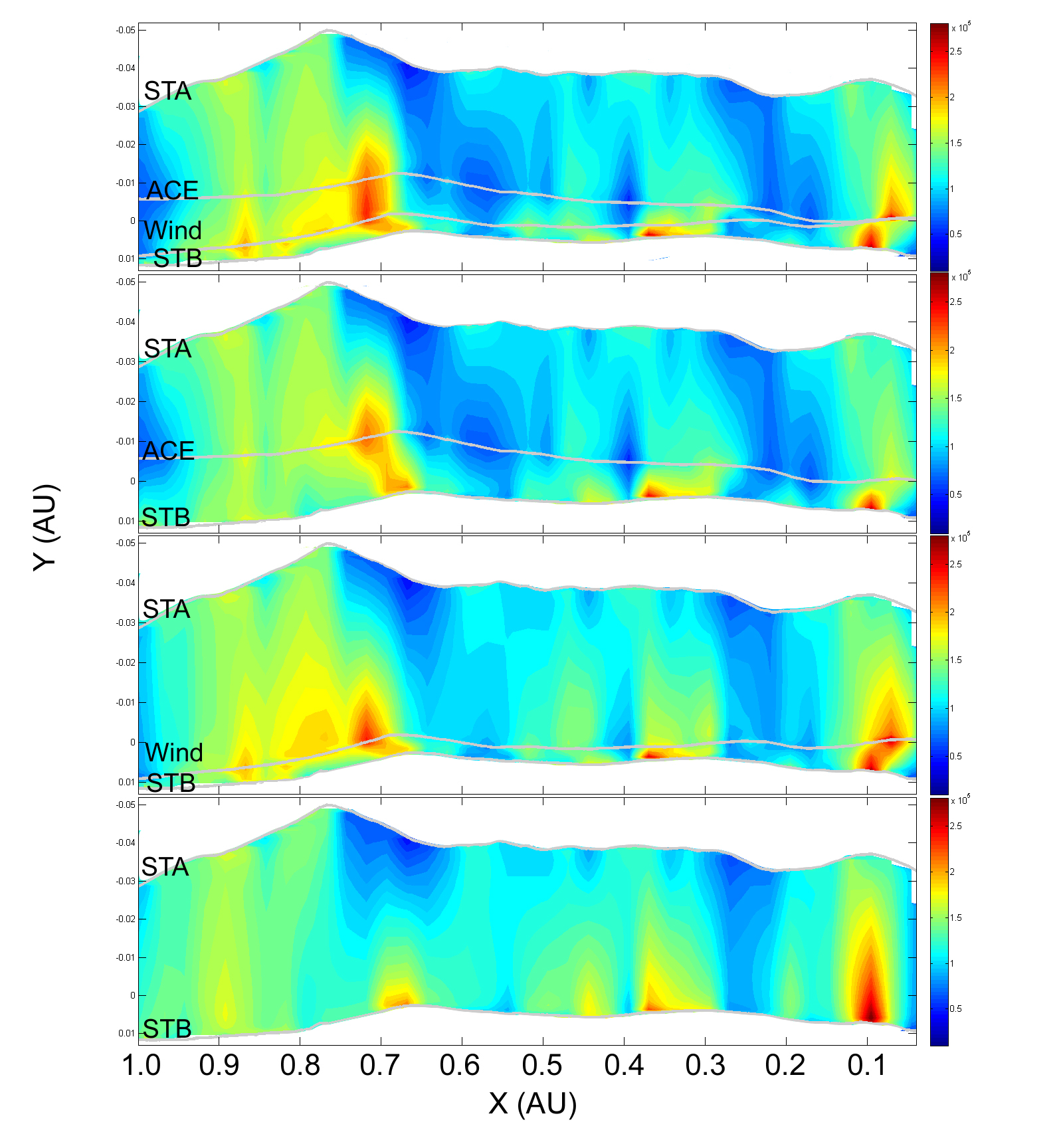}
\caption{Contour maps showing the Np scalar field interpolations for March 24 through March
29, 2007 using Wind, ACE, STA, and STB data. Top panel is a 4-spacecraft interpolation (STA,
ACE, Wind, and STB). The middle two panels use 3-spacecraft (STA, ACE, and STB and STA,
Wind, and STB) and the bottom panel uses 2-spacecraft (STA and STB) for the construction of
the proton density contour map. The grey lines show the spacecraft tracks across the map. }
\end{figure}

\begin{figure}
\includegraphics[width=200mm,bb=0 0 450 350]{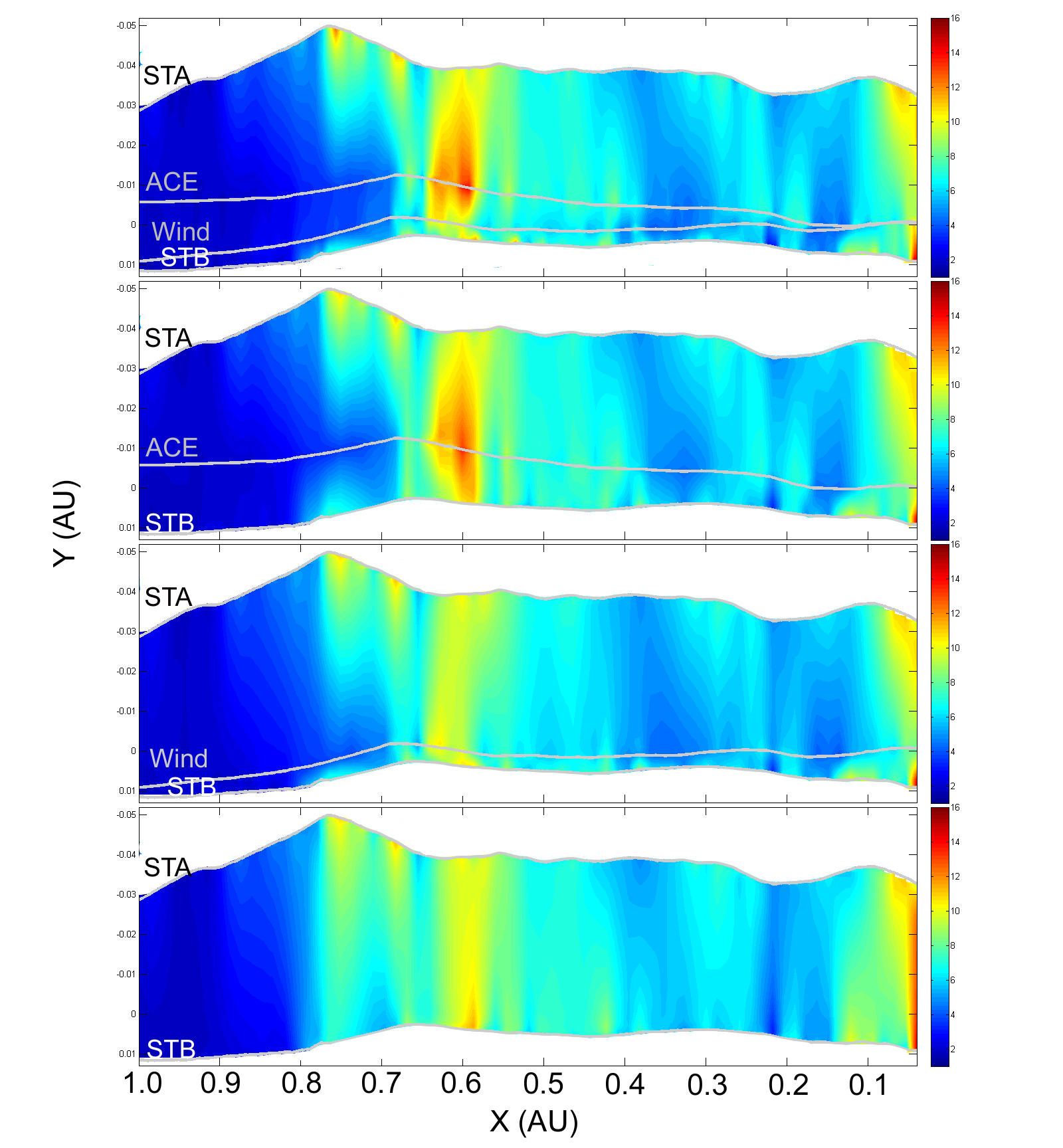}
\caption{Contour maps showing the Tp scalar field interpolations for March 24 through March
29, 2007 using Wind, ACE, STA, and STB data. Top panel is a 4-spacecraft interpolation (STA,
ACE, Wind, and STB). The middle two panels use 3-spacecraft (STA, ACE, and STB and STA,
Wind, and STB) and the bottom panel uses 2-spacecraft (STA and STB) for the construction of
the proton temperature contour map. The grey lines show the spacecraft tracks across the map. }
\end{figure}

\begin{figure}
\includegraphics[width=120mm,bb=0 0 700 700]{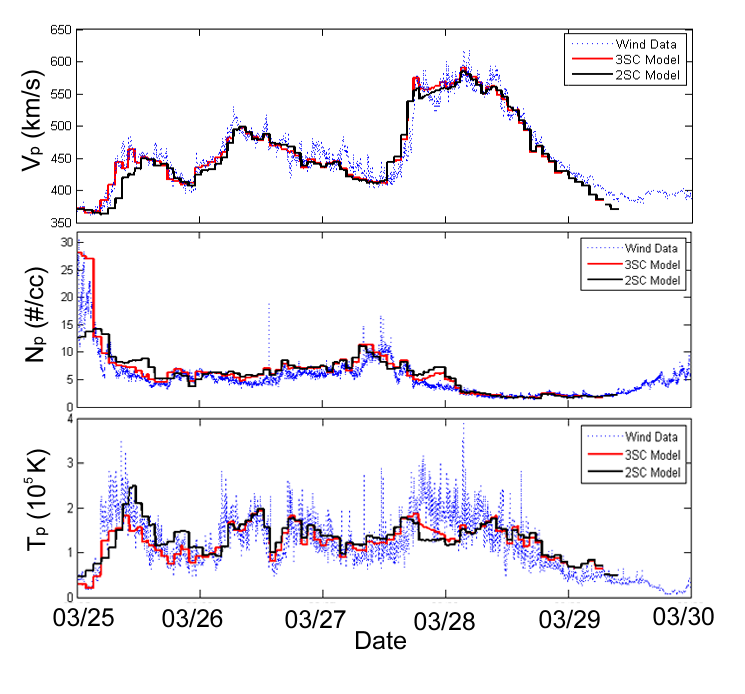}
\caption{Reconstructed time series model tracks using two- and three-spacecraft contour maps
created from Wind, ACE, STA, and STB data. Top panel shows reconstruction of Wind $V_p$ data
using the contour maps shown in the second and last panel panel of Figure 4. Middle panel
shows similar reconstruction for ACE $N_p$ data from the contour maps in the bottom two panels
of Figure A4. Bottom panel shows ACE $T_p$ data reconstructed from the contour maps in the
bottom two panels of Figure A5. In all cases the reconstructed model tracks are shown as red and
black lines and the original time series data is shown in blue. }
\end{figure}


%




%



%





%




%



%





















\end{document}